\documentclass[11pt,prd,aps,showpacs,eqsecnum,floatfix,nofootinbib,preprint,tightenlines]{revtex4}
\usepackage{graphicx}				% Use pdf, png, jpg, or eps§ with pdflatex; use eps in DVI mode
\usepackage{latexsym}
\usepackage{epsf}
\usepackage{graphicx}
\usepackage{slashed}
\usepackage{amsmath}
\usepackage{amssymb}
\usepackage{color}
\usepackage{colordvi}
\usepackage{multirow}
\newcommand{\be}{\begin{equation}}
\newcommand{\ee}{\end{equation}}
\newcommand{\bea}{\begin{eqnarray}}
\newcommand{\eea}{\end{eqnarray}}
\newcommand{\ba}{\begin{array}}
\newcommand{\ea}{\end{array}}
\newcommand{\nn}{\nonumber}

\newcommand{\hgt}{\rule[-4mm]{0mm}{10mm}}

\newcommand{\pref}[1]{(\ref{#1})}
\newcommand{\np}{N\pi}

\newcommand{\ENq}{E_{N,\vec{q}}}

\newcommand{\Epiq}{E_{\pi,{\vec{q}}}}
\newcommand{\Epip}{E_{\pi,{\vec{p}}}}

\newcommand{\GA}{G_{\rm A}}
\newcommand{\GP}{\tilde{G}_{\rm P}}

\begin{document}
\renewcommand{\thefootnote}{$*$}

\preprint{HU-EP-18/40}

\title{$N\pi$-state contamination in lattice calculations of the nucleon axial form factors}

\author{Oliver B\"ar$^{a}$} 
\affiliation{$^a$Institut f\"ur Physik,
\\Humboldt Universit\"at zu Berlin,
\\12489 Berlin, Germany\\}

\begin{abstract}
The nucleon-pion-state contribution to QCD two-point and three-point functions used in lattice calculations of the nucleon axial form factors are studied in chiral perturbation theory. For small quark masses this contribution is expected to be the dominant excited-state contamination at large time separations. To leading order in chiral perturbation theory the results depend on only two experimentally known low-energy constants and the nucleon-pion-state contribution to the form factors can be estimated. The nucleon-pion-state contribution to the axial form factor $G_{\rm A}(Q^2)$ is at the 5 percent level for a source-sink separation of 2 fm and shows almost no dependence on the momentum transfer $Q^2$. In contrast, for the induced pseudoscalar form factor $\GP(Q^2)$ the nucleon-pion-state contribution shows a rather strong dependence on $Q^2$ and leads to a 10 to 40 percent underestimation of $\GP(Q^2)$ at small momentum transfers. The ChPT results can be used to analytically remove the nucleon-pion-state contribution from lattice data. Performing this removal for lattice data generated by the PACS collaboration we find agreement with experimental data and the predictions of the pion-pole dominance model. The removal works surprisingly well even for source-sink separations as small as 1.3 fm. 
\end{abstract}

\maketitle

\renewcommand{\thefootnote}{\arabic{footnote}} \setcounter{footnote}{0}

\newpage
%========================
\section{Introduction}\label{Intro} 
%========================

Physical point simulations, i.e.\ simulations with quark masses set to their physical values, eliminate the need for a chiral extrapolation. This advantage, however, comes at a prize. Physical point simulations are numerically demanding and need substantial computer resources. The notorious signal-to-noise problem \cite{Parisi:1983ae,Lepage:1989hd} typically gets worse the lighter the pion mass is, thus the euclidean time separations in correlation functions are restricted to rather small values.\footnote{For a recently proposed method to overcome the signal-to-noise problem see Refs.\ \cite{Ce:2016idq,Ce:2016ajy}.}  At the same time the excited-state contamination due to multi-particle states involving pions grows because the energy gap to the ground state shrinks with lighter pion masses. 

The multi-particle-state contamination involving light pions can be studied using chiral perturbation theory (ChPT) \cite{Tiburzi:2009zp,Bar:2012ce,Tiburzi:2015tta}. Phenomenologically relevant is the impact of two-particle nucleon-pion ($N\pi$) states on nucleon observables, and leading order (LO) results for the nucleon mass \cite{Bar:2015zwa}, the nucleon axial, scalar and tensor charges \cite{Bar:2016uoj} as well as for various first moments of parton distribution functions \cite{Bar:2016jof} have been calculated recently.\footnote{Reviews covering these results are given in \cite{Bar:2017kxh,Bar:2017gqh}.} In case of the nucleon mass also the three-particle $N\pi\pi$-state contribution is known and found to be negligible in practice \cite{Bar:2018wco}. 

Here we present results of an analogous calculation for the $N\pi$ contamination in the nucleon axial form factors $G_{\rm A}(Q^2)$ and $\GP(Q^2)$.\footnote{Preliminary results were already reported in \cite{Bar:2018akl}.} In case of the axial form factor $G_{\rm A}(Q^2)$ one naively expects an $N\pi$ contamination similar to the one in the axial charge $g_{\rm A} = G_{\rm A}(0)$. More interesting is the induced pseudoscalar form factor $\GP(Q^2)$. Lattice calculations typically find a large excited-state contamination in this form factor, and the momentum transfer dependence expected from the pion-pole dominance (ppd) model is not reproduced by the lattice data. In this case one may suspect the source to be a low-lying $N\pi$ state, since in ChPT there contribute tree-level diagrams where the axial vector current directly creates or annihilates a pion that is absorbed or ejected at either sink or source in the three-point (3-pt) function. This potentially large contribution vanishes for zero momentum transfer, thus it is absent in the calculation of the $N\pi$ contribution to the axial charge $g_{\rm A}$ in \cite{Bar:2016uoj}. 

The results of this paper confirm these expectations. The $N\pi$-state contamination in $G_{\rm A}(Q^2)$ is essentially identical to the one in $g_{\rm A}$, leading to approximately a 5\% overestimation of the form factor by the plateau estimates at a source-sink separation of 2 fm. At the same time we find an underestimation of about 10\% to 40\% for $\GP(Q^2)$, depending strongly on the momentum transfer. The smaller $Q^2$ the larger the deviation of the plateau estimate from the true form factor. Comparing the ChPT results with recent lattice data from the PACS collaboration \cite{Ishikawa:2018rew} we observe that the $N\pi$-state contamination in the lattice plateau data reproduces a softening of the anticipated ppd behaviour at small $Q^2$, just as it has been observed in many lattice calculations so far. 

Large excited-state effects have also been observed in the 3-pt function involving the time component $A_4(x)$ of the axial vector current. In fact, the excited-state contamination is usually too large for this correlation function to be useful in the determination of the axial form factors (see Refs.\ \cite{Capitani:2017qpc,Alexandrou:2017hac}, for example). Moreover, this correlation function exhibits an unusual dependence on the operator insertion time: Instead of approaching a constant plateau value one observes an almost linear dependence \cite{Bali:2018qus}.

Also these features are qualitatively explained by the ChPT results in this paper. Expanded in inverse powers of the nucleon mass $M_N$ the ground state contribution is found to be $1/M_N$ suppressed compared to the $N\pi$-state contribution so that the latter dominates the 3-pt function. In addition, a relative sign in two terms of the $N\pi$-state contribution leads qualitatively to a dependence on the operator insertion time as it is observed in the lattice data in \cite{Bali:2018qus}.

The ChPT calculation here is analogous to the one for the axial charge in \cite{Bar:2016uoj}. The main difference is a different kinematic setup to accommodate a non-vanishing momentum transfer, as explained in section \ref{sec:ff}. While the calculation of the relevant Feynman diagrams is straightforward (sections \ref{sec:excitedstates} and \ref{sec:ffchpt}), it is more difficult to assess the range of applicability of the results, i.e.\ to estimate the minimal time separations that are necessary to apply the ChPT results (section \ref{sec:impact}). We will argue that this can be largely different for the two form factors due to different systematics in $G_{\rm A}(Q^2)$ and $\GP(Q^2)$. Eventually this can only be judged by comparing the ChPT results with lattice data, as it is done in section \ref{sec:confr}. Our main conclusions are given in section \ref{sect:concl}.

%========================
\section{The axial form factors}\label{sec:ff} 
%========================

\subsection{Basic definitions}
We consider QCD with degenerate quark masses for the light up and down quark. The spatial volume is assumed to be finite with spatial extent $L$ and periodic boundary conditions are imposed for all spatial directions. We work in euclidean space time and the time extent is taken infinite, for simplicity.

We are interested in the matrix element of the local iso-vector axial vector  current 
between single-nucleon states of definite momentum and spin,
\begin{equation}
\langle N(p',s')|A_{\mu}^a(0)|N(p,s)\rangle = \bar{u}(p',s')\left(\gamma_{\mu}\gamma_5 G_{\rm A}(Q^2) - i \gamma_5\frac{Q_{\mu}}{2M_N}\GP(Q^2)\right)\frac{\sigma^a}{2}u(p,s)\,.
\end{equation}
The right hand side shows the decomposition of the matrix element in two form factors, the axial form factor $G_{\rm A}(Q^2)$ and the induced pseudoscalar form factor $\GP(Q^2)$. $u(p,s)$ is an iso-doublet Dirac spinor with momentum $p$ and spin $s$, and the four-momentum transfer $Q_{\mu}$ reads
\begin{equation}
Q_{\mu}=(i E_{N,\vec{p}^{\,\prime}} - i E_{N,\vec{p}},\vec{q})\,,\qquad \vec{q} = \vec{p}^{\,\prime}-\vec{p}\,.
\end{equation}
Here $E_{N,\vec{p}}=\sqrt{|\vec{p}|^{\,2} +M_N^2}$ denotes the  energy of a nucleon with spatial momentum $\vec{p}$. Note that in euclidean lattice QCD the form factors are computed for space-like momentum transfers with $Q^2=(\vec{p}^{\,\prime}-\vec{p})^2 - (E_{\vec{p}^{\,\prime}} - E_{\vec{p}})^2>0$.

\subsection{Correlation functions}

The standard procedure to compute the form factors in lattice QCD is based on evaluating various 2-pt and 3-pt functions. Explicitly, the nucleon 2-pt function is given by\footnote{We always use the continuum formulation for all expressions even if we explicitly refer to quantities computed in lattice QCD.}
\begin{equation}\label{Def2ptfunc}
C_2(\vec{p},t)= \int d^3x \,e^{i\vec{p}\vec{x}}\, \Gamma_{\beta\alpha}\langle N_{\alpha}(\vec{x},t)\overline{N}_{\beta}(0,0)\rangle \,.
\end{equation}
$N,\overline{N}$ denote interpolating fields of the nucleon. Although arbitrary to a large extent we assume them to be given by the standard 3-quark operators (either pointlike or smeared) that have been mapped to ChPT, see next section.
The matrix $\Gamma$ acts on spinor space and reads
\begin{equation}\label{DefProjGamma}
\Gamma=\frac{1+\gamma_4}{4}(1+i \gamma_5\gamma_3)\,.
\end{equation}
This definition corresponds to the one employed in \cite{Alexandrou:2017hac} by the ETM collaboration, but differs by a factor 1/2 from the one used in \cite{Capitani:2017qpc}, for example. This trivial difference in normalization is irrelevant for the results of this paper.

The nucleon 3-pt function is computed with some particular kinematics: The nucleon at the sink is chosen to be at rest, $\vec{p}^{\,\prime}=0$, which implies $\vec{p}=-\vec{q}$. The expression for the momentum transfer simplifies slightly in this case,
\begin{equation}\label{DefQsqr}
Q^2=\vec{p}^{\,2} - (M_N-E_{\vec{p}})^2\,.
\end{equation}
In addition, we fix the isospin component of the axial vector current to $a=3$, so the nucleon 3-pt function we consider is given by
\begin{equation}
C_{3,{\mu}}(\vec{q},t,t')=\int d^3x\int d^3y \,e^{i\vec{q}\vec{y}}\, \Gamma_{\beta\alpha}\langle N_{\alpha}(\vec{x},t) A_{\mu}^3(\vec{y},t')\overline{N}_{\beta}(0,0)\rangle\,.
\end{equation}
The euclidean times $t$ and $t'$ denote the source-sink separation and the operator insertion time, respectively.
With the 2-pt and 3-pt functions we define the generalised ratio 
\begin{equation}\label{DefRatio}
R_{\mu}(\vec{q},t,t') =\frac{C_{3,{\mu}}(\vec{q},t,t')}{C_2(0,t)}\sqrt{\frac{C_2(\vec{q},t-t')}{C_2(0,t-t')}\frac{C_2(\vec{0},t)}{C_2(\vec{q},t)}\frac{C_2(\vec{0},t')}{C_2(\vec{q},t')}}\,.
\end{equation}
This ratio is defined in such a way that, in the asymptotic limit $t,t', t-t'\rightarrow \infty$, it converges to constant asymptotic values  $\Pi_{{\mu}}(\vec{q})$. These are related to the form factors according to ($k=1,2,3$)
\begin{eqnarray}
R_{{k}}(\vec{q},t,t') &\rightarrow &\Pi_{{k}}(\vec{q}) = \frac{i}{\sqrt{2E_{N,\vec{q}}(M_N+ E_{N,\vec{q}})}}\left( (M_N+E_{N,\vec{q}})G_{\rm A}(Q^2) \delta_{3k}-\frac{\GP(Q^2)}{2M_N} q_3q_k\right),\label{AsympValueR33}\\
R_{{4}}(\vec{q},t,t') &\rightarrow &\Pi_4(\vec{q}) = \frac{q_3}{\sqrt{2E_{N,\vec{q}}(M_N+ E_{N,\vec{q}})}}\left(G_{\rm A}(Q^2)+\frac{M_N-E_{N,\vec{q}}}{2M_N}\GP(Q^2)\right)\label{AsympValueR30}.
\end{eqnarray}

\subsection{Extracting the form factors}\label{ssect:ExtractFF}

In the results for the asymptotic ratios in eq.\ \pref{AsympValueR33}, \pref{AsympValueR30}  the two form factors enter linearly and can be extracted by solving a simple linear system. To be specific suppose we take two 3-momenta $\vec{q}_1,\vec{q}_2$ with $q_1^2=q_2^2=q^2$, thus both correspond to the same four momentum transfer $Q^2$. Eq.\ \pref{AsympValueR33}, taken twice with two spatial indices $k_1,k_2$, can be compactly written as ${\bf  \Pi = M {G}}$ with the two vectors
\begin{equation}
{\bf \Pi}
=
\left(
\begin{array}{c}
\Pi_{k_1}(\vec{q}_1)    \\
\Pi_{k_2}(\vec{q}_2) 
\end{array}
\right)\,,\qquad 
{\bf {G}}
=
\left(
\begin{array}{c}
 \GA(Q^2)    \\
 \GP(Q^2)
\end{array}
\right)\,,
\end{equation}
and the matrix
\begin{equation}\label{DefMmatrix}
{\bf M} = 
\left(
\begin{array}{cc}
c(q^2) \delta_{3,k_1}  &   d(q^2) q_{1,3}q_{1,k_1}  \\
c(q^2) \delta_{3,k_2}  &   d(q^2) q_{2,3}q_{2,k_2}      
\end{array}
\right)\,,
\end{equation}
where the functions $c(q^2), d(q^2)$ are easily read off from eq.\ \pref{AsympValueR33}. For notational simplicity we have suppressed  the dependence on the two individual momenta and components, i.e.\ we have abbreviated ${\bf \Pi}={\bf \Pi}(k_1,\vec{q}_1;k_2,\vec{q}_2)$, ${\bf M}={\bf M}(k_1,\vec{q}_1;k_2,\vec{q}_2)$ and ${\bf {G}}={\bf {G}}(Q^2)$. If eq.\ \pref{AsympValueR30} for the temporal component is part of the linear system the explicit form for ${\bf M}$ is different.  Provided the spatial momenta $\vec{q}_1,\vec{q}_2$ and the components $k_1,k_2$ are appropriately chosen the linear system has a unique solution, given as 
\begin{equation}\label{LinSystemSol}
{\bf G}= {\bf M}^{-1}{\bf \Pi} \,.
\end{equation}
in terms of the inverse of ${\bf M}$.

For our assumed setup with a finite spatial volume and periodic boundary conditions the spatial momenta are discrete,
\begin{equation}\label{pimomenta}
\vec{q}=q_L\vec{n}_q\,,\qquad q_L\,=\,\frac{2\pi}{L}\,,
\end{equation}
with the vector $\vec{n}_q$ having integer valued components. The absolute value of the momentum can be labelled by the integer $n_q$, defined according to 
\begin{equation}\label{Defnq}
q =q_L\sqrt{n_q}\,.
\end{equation}
In a finite volume only a small number of independent momenta are available as candidates for small $\vec{q}_1,\vec{q}_2$. 
All these linear systems are equivalent and yield the same results for the form factors. In practice one typically makes use of this by constructing an overdetermined linear system for the two unknown form factors, which is subsequently solved by minimizing a suitably defined least-squares function \cite{Capitani:2017qpc,Alexandrou:2017hac}.

%========================
\section{Excited-state analysis}\label{sec:excitedstates} 
%========================

\subsection{Preliminaries}

The extraction of the form factors along the lines sketched in the previous section hinges on the asymptotic values of the ratios $R_{\mu}(\vec{q},t,t')$ once all time separations $t,t'$ and $t-t'$ are taken to infinity.  In actual lattice simulations the time separations are finite and far from being asymptotically large. In that case, the 2-pt and 3-pt functions not only contain the contributions of the lowest-lying single nucleon states, but also of excited states with the same quantum numbers as the nucleon. This excited-state contamination also enters the form factors if the linear system \pref{LinSystemSol} is solved with $R_{{k}}(\vec{q},t,t')$  instead of $ \Pi_{{k}}(\vec{q})$. In other words we obtain {\rm effective} form factors $G^{\rm eff}_{\rm A}(Q^2,t,t')\,, {\tilde G}^{\rm eff}_{\rm P}(Q^2,t,t')$ including an excited-state contamination instead of the form factors we are interested in. Quite generally we expect the effective form factors to be of the form 
\begin{eqnarray}
G^{\rm eff}_{\rm A}(Q^2,t,t')\, = \,G_{\rm A}(Q^2)\bigg[ 1 + \Delta G_{\rm A}(Q^2,t,t')\bigg],\label{EffGA}\\
\GP^{\rm eff}(Q^2,t,t')\, = \,\GP(Q^2)\bigg[ 1 + \Delta \GP(Q^2,t,t')\bigg],\label{EffGP}
\end{eqnarray}
with excited state contributions $\Delta G_{\rm A}(Q^2,t,t'),\, \Delta \GP(Q^2,t,t')$ that vanish for $t,t',t-t'\rightarrow \infty$. 

For pion masses as small as in Nature one can expect two-particle $N\pi$ states with back-to-back momenta to cause the dominant excited-state contamination for large but finite time separations. This expectation rests on the naive observation that the energy gaps between the $N\pi$ states and the single nucleon ground state are smaller than those one expects from true resonance states like the Roper resonance. This not only requires small pion masses but also sufficiently large volumes such that the finite spatial momenta \pref{pimomenta} imply small energies for the lowest-lying $N\pi$ states. Volumes with $M_{\pi}L\simeq 4$, often used in lattice simulations, already fulfill this criterion \cite{Bar:2017kxh}.

In this section we derive formulae that capture the $N\pi$-state contamination in the 2-pt and 3-pt functions, the ratio $R_{\mu}$ and eventually in the effective form factors. In these expression the $N\pi$-state contamination is parameterized in terms of coefficients stemming from ratios of various matrix elements with $N\pi$ states as initial and/or final states. In the next section ChPT will be used to compute these coefficients, making the following results useful in practice.

\subsection{Correlation functions}
We start with the 2-pt function $C_2(\vec{q},t)$ defined in eq.\ \pref{Def2ptfunc}. Performing the standard spectral decomposition, projected to momentum $ \vec{q}$, the 2-pt function is a sum of various contributions,
\begin{eqnarray}\label{2ptDecomp}
C_2(\vec{q},t) & = & C^N_2(\vec{q},t) + C^{\np}_2(\vec{q},t)+\ldots
\end{eqnarray}
The first two terms on the right hand side refer to the single nucleon (SN) and the $N\pi$ contributions. The ellipsis refers to remaining contributions which we assume to be small and negligible in the following. The SN contribution is given by
\begin{equation}\label{SNcontr}
C^N_2(\vec{q},t)=\frac{1}{2E_{N,\vec{q}}}\;|\langle 0|N(0)|N(-\vec{q})\rangle|^2 e^{-E_{N,-\vec{q}}\, | t |} \,.
\end{equation}
Here $|N(-\vec{q})\rangle$ denotes the state for a moving nucleon with momentum $-\vec{q}$. The interpolating field $N(0)$ also excites $N\pi$  states with the same quantum numbers as the nucleon, thus we obtain the non-vanishing $N\pi$ contribution 
\begin{eqnarray}\label{Npicontr}
C^{\np}_2(t)&=&\frac{1}{L^3}\;\sum_{\vec{p}}\frac{1}{4E_{N,\vec{r}} E_{\pi,\vec{p}}}\,
|\langle 0|N(0)|N(\vec{r}) \pi(\vec{p})\rangle|^2 e^{-E_{\rm tot}|t|}\,.
\end{eqnarray}
The sum runs over all pion momenta that are compatible with the periodic boundary conditions, and the nucleon momentum is fixed to $\vec{r}=-\vec{q}-\vec{p}$. $E_{\rm tot}$ is the total energy of the $N\pi$ state. For weakly interacting pions $E_{\rm tot}$ equals approximately the sum $E_{N,\vec{r}}+ E_{\pi,\vec{p}}$ of the individual energies of the nucleon and the pion. 

Since the leading SN contribution is nonzero we can rewrite eq.\ \pref{2ptDecomp} as
\begin{eqnarray}
C_2(\vec{q},t) & =& C^N_2(\vec{q},t)\left\{1+ \sum_{\vec{p}} d(\vec{q},\vec{p})e^{-\Delta E(\vec{q},\vec{p}) t}\right\}\,.\label{DefC2Npcontr}
\end{eqnarray}
The coefficient $d(\vec{q},\vec{p})$ is essentially the ratio of the matrix elements in eqs.\ \pref{Npicontr} and \pref{SNcontr}, and the energy gap $\Delta E(\vec{q},\vec{p})$ reads
\begin{equation}\label{Egap2pt}
\Delta E(\vec{q},\vec{p}) = E_{\pi,\vec{p}} + E_{N,\vec{q}+\vec{p}} - \ENq\,.
\end{equation}
Here, as mentioned before, we have ignored the interaction energy. In the next section we compute the 2-pt function in ChPT, and to LO we will recover the result  \pref{Egap2pt} for the energy gap. Deviations due to the nucleon-pion interaction will show up at higher order in the chiral expansion.

The 2-pt function enters the generalized ratio $R_{\mu}(\vec{q},t,t')$. Introducing the short hand notation $\sqrt{\Pi C_2}$ for the square root expression in \pref{DefRatio} and expanding in powers of small quantities we obtain
\begin{equation}\label{ratio2ptGeneric}
\frac{1}{C_2(0,t)}\sqrt{\Pi C_2} = \frac{1}{C^N_2(0,t)}\sqrt{\Pi C^N_2} \left\{ 1+ \frac{1}{2} Y(\vec{q},\vec{p})\right\}\,,
\end{equation}
where the function $Y(\vec{q},\vec{p})$ contains the $\np$-state contribution,
\begin{eqnarray}
Y(\vec{q},\vec{p})&=&\sum_{\vec{p}}\Bigg(d(\vec{q},\vec{p})  \left\{ e^{-\Delta E(\vec{q},\vec{p}) (t-t')} - e^{-\Delta E(\vec{q},\vec{p}) t'} - e^{-\Delta E(\vec{q},\vec{p}) t}\right\}\nn\\
& & \phantom{\sum_{\vec{p}}} - d(0,\vec{p})  \left\{ e^{-\Delta E(\vec{0},\vec{p}) (t-t')} - e^{-\Delta E(\vec{0},\vec{p}) t'} + e^{-\Delta E(\vec{0},\vec{p}) t}\right\}\Bigg)\,.\label{DefY}
\end{eqnarray}

The excited-state analysis of the 3-pt function follows the same lines. Performing again the spectral decomposition we find, in analogy to \pref{2ptDecomp}, the result 
\begin{eqnarray}
C_{3,\mu}(\vec{q},t,t') & = & C^N_{3,\mu}(\vec{q},t,t') + C^{\np}_{3,\mu}(\vec{q},t,t')+\ldots\,.
\end{eqnarray}
As before we will ignore all but the SN and the $N\pi$ contribution in the following. The expressions analogous to eqs.\ \pref{SNcontr} and \pref{Npicontr} for $C^N_{3,\mu}(\vec{q},t,t')$ and $C^{\np}_{3,\mu}(\vec{q},t,t')$ can be worked out straightforwardly but they are slightly cumbersome.  Eventually we need the parameterization analogues to \pref{2ptDecomp}, where we write (no summation over $\mu$!)
\begin{eqnarray}
C_{3,\mu}(\vec{q},t,t') & =& C^N_{3,\mu}(\vec{q},t,t')\bigg(1+ Z_{\mu}(\vec{q},t,t')\bigg) \,,\label{DefC3Npcontr}
\end{eqnarray}
i.e.\ $Z_{\mu}$ denotes the ratio $C^{\np}_{3,\mu}(\vec{q},t,t')/C^N_{3,\mu}(\vec{q},t,t')$. Here we assume and only consider the cases where the SN contribution is non-vanishing, which puts a constraint on the possible momenta $\vec{q}$ and the index $\mu$. The generic form for $Z_{\mu}(\vec{q},t,t')$ is found as
\begin{eqnarray}
Z_{\mu}(\vec{q},t,t') & = & a_{\mu}(\vec{q}) e^{-\Delta E(0,-\vec{q}) (t-t')}+ \tilde{a}_{\mu}(\vec{q}) e^{-\Delta E(\vec{q},-\vec{q})t'} \nn \\[2ex]
& & + \sum_{\vec{p}} b_{\mu}(\vec{q},\vec{p}) e^{-\Delta E(0,\vec{p}) (t-t')}+\sum_{\vec{p}}\tilde{b}_{\mu}(\vec{q},\vec{p}) e^{-\Delta E(q,\vec{p}) t'} \nn\\
& & +  \sum_{\vec{p}} c_{\mu}(\vec{q},\vec{p}) e^{-\Delta E(0,\vec{p}) (t-t')}e^{-\Delta E(\vec{q},\vec{p}) t'}\,.\label{DefZmu}
\end{eqnarray}
with the energy gaps specified in eq.\ \pref{Egap2pt}. 
The coefficients $a_{\mu}(\vec{q}), \tilde{a}_{\mu}(\vec{q}),b_{\mu}(\vec{q},\vec{p}) ,\tilde{b}_{\mu}(\vec{q},\vec{p}), c_{\mu}(\vec{q},\vec{p})$ in \pref{DefZmu} contain ratios of matrix elements involving the nucleon interpolating fields and the axial vector current. For example, the coefficient $b_{\mu}(\vec{q},\vec{p})$ contains the matrix element $\langle N\pi |A_{\mu}^a|N \rangle$ with the $N\pi$ state as the final state.  Similarly, $\tilde{b}_{\mu}(\vec{q},\vec{p})$ contains the matrix element with the $N\pi$ state as the initial state. Together the $b_{\mu}(\vec{q},\vec{p})$ and $\tilde{b}_{\mu}(\vec{q},\vec{p})$ contribution forms the excited-to-ground-state contribution. Similarly, the $c_{\mu}(\vec{q},\vec{p})$ contribution is called the excited-to-excited-state contribution, since it involves the matrix elements with $N\pi$ states as initial and final states. As before the sums run over the momentum of the pion in the $N\pi$ state. The associated nucleon momentum is fixed by momentum conservation and the kinematic setup we have chosen.

The presence of the contributions proportional to $a_{\mu}(\vec{q})$ and $\tilde{a}_{\mu}(\vec{q})$ deserves a comment. As it stands these are also captured by the ones proportional to $b_{\mu}(\vec{q},-\vec{q})$ and $\tilde{b}_{\mu}(\vec{q},-\vec{q})$. The reason for this   separation and a precise definition of the coefficients will be given in the next section when we compute the coefficients perturbatively in ChPT.

Taking the product of \pref{DefC3Npcontr} and \pref{ratio2ptGeneric} we obtain the total result for the $N\pi$ contamination in the generalized ratios,
\begin{eqnarray}
R_{\mu}(\vec{q},t,t')&=& \Pi_{\mu}(\vec{q}) \Bigg(1+ Z_{\mu}(\vec{q},t,t') + \frac{1}{2} Y(\vec{q},t,t')\Bigg)\,,\\
& \equiv & \Pi_{\mu}(\vec{q}) \Bigg(1+ X_{\mu}(\vec{q},t,t')\Bigg)\label{NpiConttot}
\end{eqnarray}
with $\Pi_{\mu}(\vec{q})$ referring to the asymptotic values of the ratios introduced in \pref{AsympValueR33} and  \pref{AsympValueR30}. The $N\pi$ contamination $X_{\mu}(\vec{q},t,t')$ vanishes exponentially as the time separations tend to infinity, so the ratios  correctly approach their asymptotic values. 

\subsection{Effective form factors}
In practice the ratios $R_{\mu}(\vec{q},t,t')$ are at our disposal for moderately large time separations, not their asymptotic values $\Pi_{\mu}(\vec{q})$. The $N\pi$ state contribution present in these ratios modifies the solution of the linear system that we solve to extract the form factors. Instead of the true form factors ${\bf G}$ in eq.\ \pref{LinSystemSol} we get effective form factors that inherit the $N\pi$ contribution and its time dependence,
 \begin{equation}\label{Geffa}
{\bf 
 {G}^{\rm eff}}(\vec{q},t,t')= {\bf M^{-1}}\left({\bf {\Pi}}(Q^2) +{\bf \Delta {\Pi}}(\vec{q},t,t')\right)\,.
\end{equation}
where the correction term on the right hand side reads
\begin{equation}
{\bf \Delta \Pi}(\vec{q},t,t')
=
\left(
\begin{array}{c}
 \Pi_{k_1}(\vec{q}_1)  X_{k_1}(\vec{q}_1,t,t')  \\
 \Pi_{k_2}(\vec{q}_2) X_{k_2}(\vec{q}_2,t,t') 
\end{array}
\right)\,.
\end{equation}
Eqs.\ \pref{EffGA} and \pref{EffGP}, quoted at the beginning of this section, are nothing but \pref{Geffa}, slightly rewritten and with the dependence on the momentum transfer made explicit.

The main result here is that both form factors inherit the $N\pi$ state contribution of the two particular ratios that we have chosen for our linear system, and these depend on the particular combinations $k_1,\vec{q}_1$ and  $k_2,\vec{q}_2$ for the momenta and the spatial indices of the axial vector current. This is can be interesting if there are various combinations possible that can be used to extract the form factors and if some combinations show  significantly smaller $N\pi$ contaminations than others. We will study this question in section \ref{ssect:impactonPlatest}. 

%========================
\section{$N\pi$-state contribution in ChPT}\label{sec:ffchpt} 
%========================

\subsection{Preliminaries}
The correlation functions and their ratios $R_{\mu}$ defined in the previous section can be computed in ChPT, the low-energy effective theory of QCD \cite{Weinberg:1978kz,Gasser:1983yg,Gasser:1984gg}. For sufficiently large times $t,t',$ pion physics will dominate the correlation functions and ChPT is expected to provide good estimates for them. In particular, we can obtain ChPT results for the coefficients $d(\vec{q},\vec{p})$ in \pref{2ptDecomp} and $a_{\mu}(\vec{q}), \tilde{a}_{\mu}(\vec{q}),b_{\mu}(\vec{q},\vec{p}) ,\tilde{b}_{\mu}(\vec{q},\vec{p}), c_{\mu}(\vec{q},\vec{p})$ in \pref{DefZmu}. 

Such calculations have been performed and described before, see \cite{Bar:2015zwa,Tiburzi:2015tta,Bar:2016uoj,Bar:2016jof}. Ref.\ \cite{Bar:2016uoj} reports a ChPT calculation for the  
the axial vector 3-pt function for vanishing momentum transfer and the $N\pi$-state contamination in lattice calculations of the axial charge $g_{A} =\GA(0)$. The calculation presented here is completely analogous, the main difference is the different kinematics in the correlation functions and the extraction of the form factors for non-vanishing $Q^2$. The ChPT setup with the chiral expressions for the axial vector current and the nucleon interpolating fields is independent of the kinematics and the same as in Ref.\ \cite{Bar:2016uoj}. For completeness and the reader's convenience the Feynman rules are summarized in appendix \ref{appFeynmanRules}. For details, however, the reader is referred to \cite{Bar:2016uoj} and the reviews \cite{Bar:2017kxh,Bar:2017gqh}.

The calculation is done in covariant ChPT \cite{Gasser:1987rb,Becher:1999he} to LO. At this order the results for the various coefficients depend on two LO low-energy coefficients (LECs) only, the pion decay constant and the axial charge. Since these are known phenomenologically very precisely the LO ChPT results are very predictive. In particular, they do not depend on the LECs associated with the nucleon interpolating fields \cite{Wein:2011ix}, because these drop out at LO.  
 
%========================
\subsection{Form factors - Single nucleon contribution}\label{ssec:SN} 
%========================

%Figure SN diagrams
\begin{figure}[t]
\begin{center}
\includegraphics[scale=1.0]{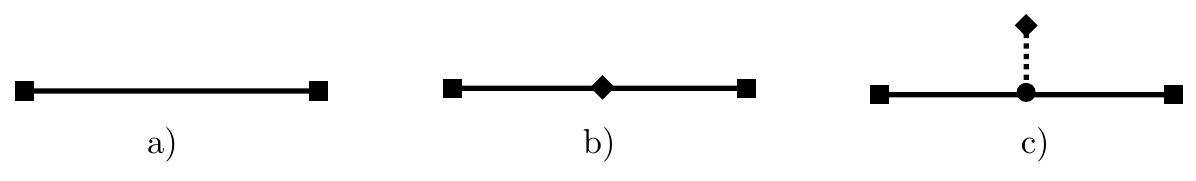}\\
\caption{Feynman diagrams for the leading SN contribution in the  2-pt function (diagram a) and the 3-pt function (diagrams b) and c). Squares represent the nucleon nucleon interpolating fields at times $t$ and $0$. The diamond stands for the axial vector current at insertion time $t'$.  The circle represents a vertex insertion at an intermediate space time point, and an integration over this point is implicitly assumed. The solid (dashed) lines represent a nucleon (pion) propagator in the time-momentum representation, see appendix \ref{appFeynmanRules}.}
\label{fig:diagsSN}
\end{center}
\end{figure}
% End figure

Figure \ \ref{fig:diagsSN} shows the Feynman diagrams that provide the leading SN contribution to the 2-pt and 3-pt functions and, consequently, the ratios and the form factors. The calculation is simple and essentially establishes that we have properly matched ChPT to QCD.

The result for the 2-pt diagram reads (we always assume $t> 0$)
\begin{equation}\label{C2qSN}
C_2(\vec{q},t) = \frac{\ENq +M_N}{2 \ENq} e^{-\ENq  t}\,.
\end{equation}
For $\vec{q}=0$ we recover the result derived in Ref.\ \cite{Bar:2015zwa}. 
The SN contribution to the 3-pt function is the sum of the results for diagrams b) and c) in fig.\ \ref{fig:diagsSN}:
\begin{eqnarray}
C_{3,k}(\vec{q},t,t') & = & \frac{ig_A}{2\ENq}  \left((\ENq +M_N)\delta_{3,k}+ \frac{2M_N}{(\ENq-M_N)^2-\Epiq^2} q_3q_k\right)e^{-M_N(t-t')} e^{-\ENq t'}\label{C3SNk}\\
C_{3,{4}} (\vec{q},t,t')& = & \frac{q_3g_A}{2\ENq} \left( 1 + \frac{2(\ENq-M_N)M_N}{(\ENq-M_N)^2-\Epiq^2}   \right)\label{C3SN0}e^{-M_N(t-t')} e^{-\ENq t'}
\end{eqnarray}
With these results it is straightforward to compute the ratios $R_{\mu}$ and extract the two form factors as described in section \ref{ssect:ExtractFF}, and we obtain
\begin{eqnarray}\label{FFSN}
&&\GA(Q^2)\,  =  \, g_A\,,\quad
\GP(Q^2) \, = \,  4 g_A\frac{M_N^2}{Q^2+M_{\pi}^2}\,.
\end{eqnarray}
The result for the axial form factor agrees with the one in Ref.\ \cite{Bar:2016uoj} for vanishing momentum transfer. The result for the induced pseudoscalar form factor agrees with the one in Ref.\ \cite{Schindler:2006it}.

%========================
\subsection{$N\pi$-state contribution - general remarks}\label{ssec:coeff} 
%========================

The $\np$-state contribution to the 2-pt function and the axial vector 3-pt function with $\vec{q}=0$ has already been computed in the covariant formulation of BChPT, see Ref.\ \cite{Bar:2015zwa,Bar:2016uoj}. The results for the coefficients are sufficiently compact in the full covariant form. However, already the 2-pt function is fairly cumbersome once we allow  $\vec{q}\neq0$. The expressions simplify significantly if we perform the non-relativistic (NR) expansion, i.e.\ if we expand the nucleon energy according to 
\begin{equation}\label{NRexpansion}
\ENq = M_N+\frac{\vec{q}^{\,2}}{2M_N} 
\end{equation}
and keep only the first two terms. For practical uses this should be sufficient.

The NR expansion is slightly different for the spatial components and $\mu = 4$. The reason is that the SN contribution to the 3-pt function has a different non-relativistic limit for $\mu=k$ and $\mu=4$. Performing \pref{NRexpansion} in \pref{C3SN0} one finds
\begin{eqnarray}
C^N_{3,4}(\vec{q},t,t') & = &g_A \frac{M_{\pi}^2 q_3}{2 \Epiq^2 M_N}+{\rm O}\left(\frac{1}{M_N^3}\right)\,.\label{CN34}
\end{eqnarray}
Thus, it vanishes in the infinite nucleon mass limit. On the other hand, the SN contribution of the spatial components in \pref{C3SNk} 
 have a non-vanishing NR limit value.  

The coefficients we want compute in ChPT are the ones in $Z_{\mu}$, which itself is defined as the ratio $C^{N\pi}_{3,\mu}/C^N_{3,\mu}$, cf.\ eq.\ \pref{DefC3Npcontr}. We will find that the numerator in this ratio typically starts with O$(1)$ in the NR expansion. Consequently, for $\mu=4$, the  inverse power $1/M_N$ in the SN contribution shifts the NR expansion of the ratio such that powers linear in the nucleon mass appear. In that case the coefficients diverge in the infinite nucleon mass limit, simply because the single nucleon contribution vanishes in this limit while the $N\pi$ contribution tends to a non-vanishing constant. 

%========================
\subsection{$N\pi$-state contribution - the 2-pt function}
%========================

% Figure: Diagrams for nucleon pion contributions in 2pt function
\begin{figure}[t]
\begin{center}
\includegraphics[scale=1.0]{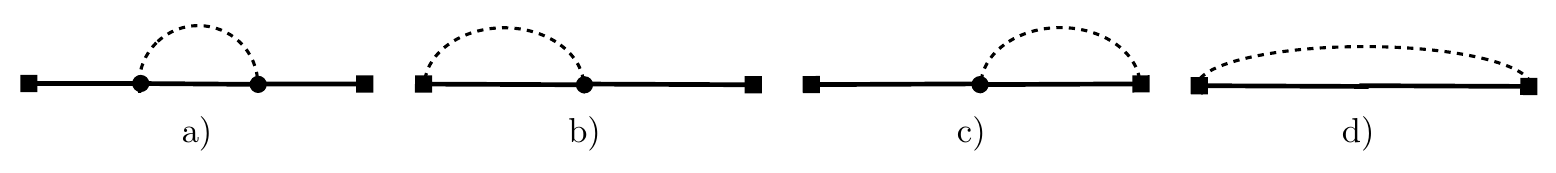}\\
\caption{Feynman diagrams for the leading $N\pi$ contribution in the 2-pt function.}
\label{fig:Npidiags2pt}
\end{center}
\end{figure}
% End figure

We introduced the coefficients $d(\vec{q},\vec{p})$ in eq.\ \pref{DefC2Npcontr}. These coefficients can be calculated by computing the four loop diagrams for the 2-pt function shown in figure \ref{fig:Npidiags2pt}. Following \cite{Bar:2015zwa,Bar:2016uoj} we separate the coefficients into an universal part and a ``reduced coefficient'',
\begin{equation}\label{DefRedCoeff}
d(\vec{q},\vec{p}) = \frac{1}{8 (fL)^2 E_{\pi,\vec{p}}L} D(\vec{q},\vec{p}). 
\end{equation}
The denominator contains the dimensionless combinations $fL$ and $E_{\pi}L$ that we expect to appear in loop diagrams.\footnote{Note that this definition differs slightly from the one used in \cite{Bar:2015zwa}. The universal factor in \pref{DefRedCoeff} does not include a factor 3 in the numerator stemming from flavor symmetry.} To the order we are working here the reduced coefficients $D(\vec{q},\vec{p})$ are dimensionless functions of the nucleon and pion energies and the axial charge, see below.

As mentioned before, we perform the NR expansion, thus we write the coefficients in the following form:
\begin{equation}\label{DefNRExp}
D(\vec{q},\vec{p}) =D^{\infty}(\vec{q},\vec{p}) + \frac{E_{\pi,\vec{p}}}{M_N}D^{\rm corr}(\vec{q},\vec{p})\,.
\end{equation}
The particular form of the $\np$-vertex in the interpolating nucleon field implies that only diagram a) contributes to the infinite nucleon mass limit  $D^{\infty}(\vec{q},\vec{p})$. Contributions to the correction $D^{\rm corr}(\vec{q},\vec{p})$ originate in diagrams a), b) and c), while d) contributes to O($1/M_N^2$) only and can be ignored. 

The calculation parallels the one for $\vec{q}=0$ done in Ref.\ \cite{Bar:2015zwa}, and the results are:
\begin{eqnarray}
D^{\infty}(\vec{q},\vec{p}) & = & 3g_A^2\frac{ p^2}{E_{\pi,\vec{p}}^2}\,,\\
D^{\rm corr}(\vec{q},\vec{p}) & = & 3g_A\frac{g_A M_{\pi}^2 (p^2+2pq)-E_{\pi,\vec{p}}^2 (p^2+pq)}{E_{\pi,\vec{p}}^4}\,.
\end{eqnarray}
The NR limit result $D^{\infty}(\vec{q},\vec{p})$ does not depend on the injected momentum $\vec{q}$ and can directly be compared with the result for $\vec{q}=0$. The correction, however, does depend on $\vec{q}$ in form of the scalar product $pq=\vec{p}\cdot\vec{q}$. Setting this scalar product to zero we obtain 
\begin{equation}
D^{\rm corr}(0,\vec{p})  = 3g_A \frac{p^2}{E^2_{\pi,\vec{p}}} \left(g_A \frac{M_{\pi}^2}{E_{\pi,\vec{p}}^2}-1\right).
\end{equation}
The result for $\vec{q}=0$ agree with the one in Ref.\ \cite{Bar:2015zwa}.

%========================
\subsection{$N\pi$-state contribution - the 3-pt function}
%========================

% Figure: Diagrams for nucleon pion contributions
\begin{figure}[t]
\begin{center}
\includegraphics[scale=1.0]{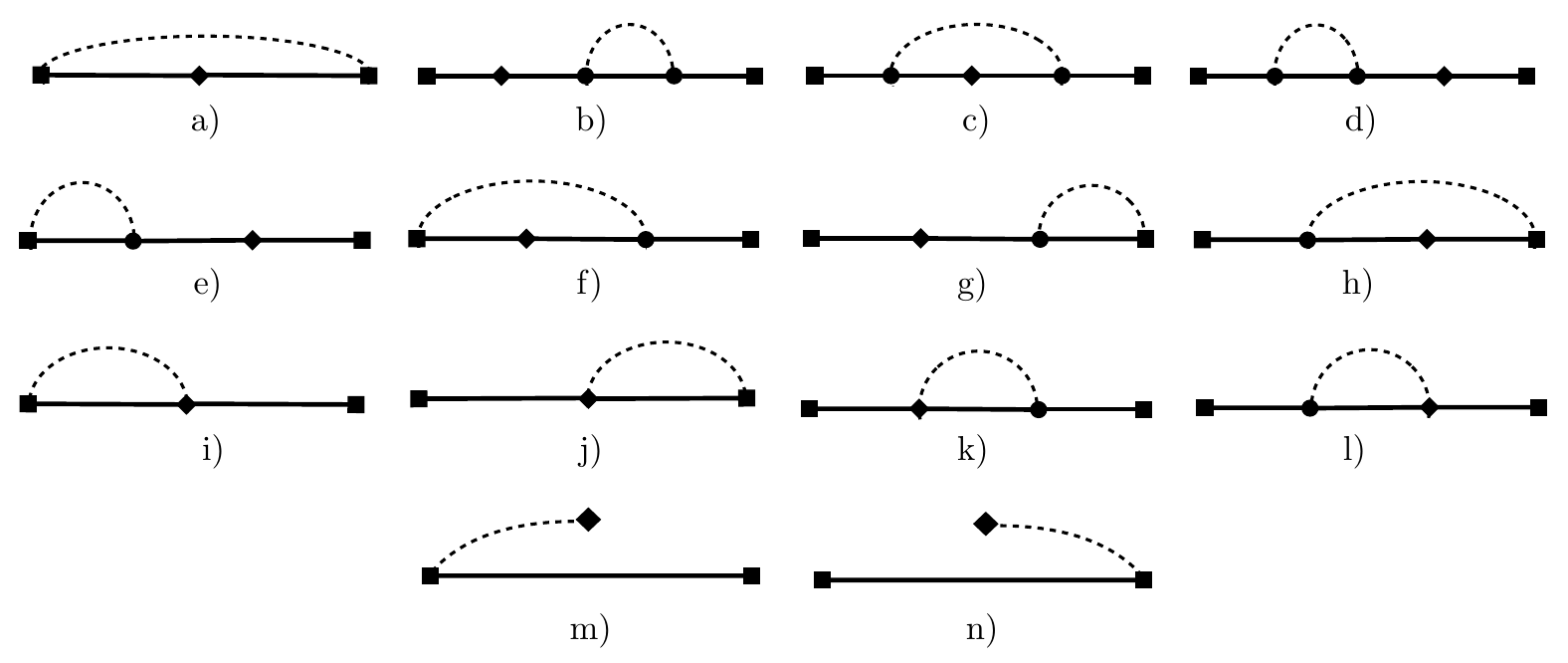}\\
\caption{Feynman diagrams for the leading $N\pi$ contribution in the 3-pt functions.}
\label{fig:Npidiags}
\end{center}
\end{figure}
% End figure

Figure \ref{fig:Npidiags} shows the diagrams with a nonzero $N\pi$-state contribution to the 3-pt functions. The first twelve loop diagrams are the same as the ones calculated in \cite{Bar:2016uoj} for vanishing momentum transfer. For $\vec{q}\neq0$ the two remaining tree diagrams and diagram c) in fig.\ \ref{fig:diagsSN} also contribute.  

For later reference it will be useful to keep the contributions from the loop and the tree diagrams separate. The coefficients $b_{\mu}(\vec{q},\vec{p}) ,\tilde{b}_{\mu}(\vec{q},\vec{p}), c_{\mu}(\vec{q},\vec{p})$ in \pref{DefZmu} capture the $N\pi$ contribution of the loop diagrams, while the tree diagram contribution is given by $a_{\mu}(\vec{q}), \tilde{a}_{\mu}(\vec{q})$. For the spatial components of the latter we introduce the NR expansion according to 
\begin{equation}\label{DefaAndat}
a_{k}(\vec{q}) = a^{\infty}_{k}(\vec{q}) +\frac{E_{\pi,\vec{q}}}{M_N} a^{\rm corr}_{k}(\vec{q})\,,\qquad \tilde{a}_{k}(\vec{q}) = \tilde{a}^{\infty}_{k}(\vec{q}) +\frac{E_{\pi,\vec{q}}}{M_N} \tilde{a}^{\rm corr}_{k}(\vec{q})\,.
\end{equation}
As explained before, the NR expansion for $\mu=4$ is ``shifted'' due to the normalization with $C^N_{3,4}(\vec{q},t,t')$ given in \pref{CN34}. Thus we introduce
\begin{equation}
a_4(\vec{q}) = \frac{M_N}{E_{\pi,\vec{q}}} a^{\infty}_{4}(\vec{q}) + a^{\rm corr}_{4}(\vec{q})\,,\qquad \tilde{a}_{4}(\vec{q}) = \frac{M_N}{E_{\pi,\vec{q}}}\tilde{a}^{\infty}_{4}(\vec{q}) + \tilde{a}^{\rm corr}_{4}(\vec{q}).
\end{equation}
Note that the tree level diagrams do not lead to an inverse power of the volume $L^3$, thus there is no factor analogous to the universal factor in \pref{DefRedCoeff}. Note also that these coefficients depend on one momentum only, the momentum transfer $\vec{q}$. 
The results for these coefficients are as follows: 
\begin{eqnarray}\label{resLimaAndat}
a^{\infty}_{k=1,2}(\vec{q})& = & - \frac{1}{2}\,,\qquad a^{\infty}_{k=3}(\vec{q})\,=\, \frac{1}{2}\frac{q_3^2}{E_{\pi,\vec{q}}^2-q^2_3} \,=\,\frac{q_3^2}{2(q^2_1+q_2^2 +M_{\pi}^2)}\,,
\end{eqnarray}
and the same results for the coefficient $\tilde{a}^{\infty}_{k}(\vec{q})$,
\begin{eqnarray}\label{symakinfty}
\tilde{a}^{\infty}_{k}(\vec{q}) &=&a^{\infty}_{k}(\vec{q})\,,\quad k\,=\,1,2,3\,.
\end{eqnarray}
For $k=1,2$ the coefficient is constant, for $k=3$ it can vanish if the third momentum component $q_3$ is zero.
The simplicity of these results is a consequence of the fact that the NR limit values stem from diagram c) in fig.\ \ref{fig:diagsSN} only. The remaining two involve the $N\pi$ vertex in the interpolating fields and are therefore expected not to contribute to the NR limit values. This is explicitly found in the calculation. 

For the corrections we find
\begin{eqnarray}
a^{\rm corr}_{k}(\vec{q})& = & \frac{1}{2}\left(\frac{M_{\pi}^2}{\Epiq^2}-\frac{1}{g_A}\right)a^{\infty}_{k}(\vec{q})\,,\quad k\,=\,1,2,3\,,\\
\tilde{a}^{\rm corr}_{k}(\vec{q})& = & \frac{1}{2}\left(\frac{M_{\pi}^2}{\Epiq^2}\right)\tilde{a}^{\infty}_{k}(\vec{q}),\quad k\,=\,1,2,3\,.\label{atildecorrk}
\end{eqnarray}
The results are not the same for ${a}^{\rm corr}_{\mu}(\vec{q})$ and $\tilde{a}^{\rm corr}_{\mu}(\vec{q})$. One reason is that diagram  n) in fig.\ \ref{fig:Npidiags} vanishes, while diagram m) contributes with the term proportional to $1/g_A$ in ${a}^{\rm corr}_{\mu}(\vec{q})$. This is a property of our particular kinematic setup with a vanishing momentum $\vec{p}^{\,\prime}$ for the nucleon in the final state.

Note that the correction coefficients vanish if the leading ones are zero. In particular, the coefficients $a_{3}(\vec{q})$ and $\tilde{a}_{3}(\vec{q})$ vanish for $\vec{q}=0$. Therefore, the tree diagrams do not contribute to $R_{3}(\vec{q}=0,t,t')$  in \pref{NpiConttot}, the ratio necessary for  the calculation of the axial charge $g_A$ \cite{Bar:2016uoj}.

For typical pion energies the correction coefficients result in a small O($1/M_N$) contribution. For example, for pion energies of 280 MeV the correction coefficient is about a quarter of the leading one. Taking into account the suppression factor $\ENq/M_N$ we roughly obtain a  10\% correction due to the correction coefficients.

The results for $\mu=4$ are as follows:
\begin{eqnarray}
a^{\infty}_{4}(\vec{q}) & = & -\frac{\Epiq^2}{M^2_{\pi}}\,,\qquad a^{\rm corr}_{4}(\vec{q}) \,=\,-\frac{1}{2}\left(1-\frac{\Epiq^2}{g_A M_{\pi}^2}\right)\,,\label{a4}\\
\tilde{a}^{\infty}_{4}(\vec{q}) & = & \frac{\Epiq^2}{M^2_{\pi}}\,,\qquad \tilde{a}^{\rm corr}_{4}(\vec{q}) \,=\, -\frac{1}{2}\,.\label{at4}
\end{eqnarray}
Note that $a^{\infty}_{4}(\vec{q}) = - \tilde{a}^{\infty}_{4}(\vec{q})$, in contrast to the coefficients for $\mu=k$, where \pref{symakinfty} holds. This difference will be responsible for a qualitatively different behavior of the ratio $R_4(\vec{q},t,t')$ involving the time component of the axial vector current, see section \ref{sect:ImpactA4}.

For the loop diagram contribution we define, in analogy to eq.\ \pref{DefRedCoeff}, reduced coefficients denoted by capital letters $B,\tilde{B}$ and $C$. These are expanded according to the NR expansion. For the spatial components this is equivalent to eq.\ \pref{DefNRExp}. For example, 
\begin{equation}\label{def:Bk}
B_{k}(\vec{q},\vec{p}) =B_{k}^{\infty}(\vec{q},\vec{p}) + \frac{E_{\pi,\vec{p}}}{M_N}B_{k}^{\rm corr}(\vec{q},\vec{p})\,,
\end{equation}
and analogously for $\tilde{B}_{k}, C_{k}$. In contrast, for $\mu=4$, we introduce
\begin{equation}\label{def:B4}
B_{4}(\vec{q},\vec{p}) =\frac{M_N}{E_{\pi,\vec{p}}}B_{4}^{\infty}(\vec{q},\vec{p}) + B_{4}^{\rm corr}(\vec{q},\vec{p})\,,
\end{equation}
and here too analogous expressions for the other two coefficients. The results for these coefficients are as follows. 
For the spatial components we find
\begin{eqnarray}
B_{k}^{\infty}(\vec{q},\vec{p}) &=&-2g_A^2 \frac{\Epiq^2}{\Epip^2}\frac{p_kp_3}{q_kq_3}\,,\qquad k\,=\,1,2\,,\label{resLimBk}\\
B_{3}^{\infty}(\vec{q},\vec{p}) &=&\phantom{-}2g_A^2 \frac{\Epiq^2}{\Epip^2}\frac{p^2 +p_3^2}{\Epiq^2-q_3^2}\label{resLimB3}\,,
\end{eqnarray}
together with the relation
\begin{eqnarray}
\tilde{B}_{k}^{\infty}(\vec{q},\vec{p})&=&B_{k}^{\infty}(\vec{q},\vec{p})\,,\qquad k\,=\,1,2,3\,.\label{resLimBtk}
\end{eqnarray}
For the remaining coefficient we find
\begin{eqnarray}
C_{k}^{\infty}(\vec{q},\vec{p}) &=&-B_{k}^{\infty}(\vec{q},\vec{p}) \,,\qquad k\,=\,1,2\,,\\
C_{3}^{\infty}(\vec{q},\vec{p}) &=&\phantom{-}g_A^2 \frac{\Epiq^2}{\Epip^2}\frac{p^2 -2p_3^2}{\Epiq^2-q_3^2}\,.
\end{eqnarray}
Finally, for the $\mu=4$ component the results read
\begin{eqnarray}
B_{4}^{\infty}(\vec{q},\vec{p}) &=&-8\frac{\Epiq^2 p_3}{M_{\pi}^2 q_3}\,,\\
\tilde{B}_{4}^{\infty}(\vec{q},\vec{p}) &=&B_{4}^{\infty}(\vec{q},\vec{p})\,,\\
C_4^{\infty}(\vec{q},\vec{p})  &=&0\,.
\end{eqnarray}
The results for the correction coefficients $B_{k}^{\rm corr}(\vec{q},\vec{p}),\tilde{B}_{k}^{\rm corr}(\vec{q},\vec{p})$ and $C_{k}^{\rm corr}(\vec{q},\vec{p})$ are slightly cumbersome. Since the detailed expressions reveal no additional qualitative insights they are listed in appendix \ref{app:corrcoeff}.

%========================
\section{Impact on lattice calculations}\label{sec:impact} 
%========================

\subsection{Preliminaries}\label{sect:ImpactPrelim}

To LO in ChPT the $N\pi$ contribution to the ratios $R_{\mu}$ and the effective form factors depends on a few LECs only, and these are known rather precisely from experiment. Assuming these values in the ChPT results we obtain estimates for the expected impact of the $N\pi$ contribution  in lattice QCD simulations. 

The LECs are the chiral limit values of the pion decay constant and the axial charge.  To LO it is consistent to use the experimental values for these LECs and we set  $g_A=1.27$ and $f=f_{\pi}= 93$ MeV \cite{Tanabashi:2018oca}. We can ignore the errors in these values since they are too small to be significant for the LO estimates. Since we are mainly interested in the $N\pi$ contribution in physical point simulations we fix the pion and nucleon masses to their physical values. In the following it is sufficient to use the simple estimates $M_{\pi}=140$ MeV and $M_N=940$ MeV. 

We also need to fix the size of the spatial volume, and we do this by imposing a value for $M_{\pi}L$.  In Refs.\ \cite{Bar:2016uoj,Bar:2016jof} the FV effects of the $N\pi$ contribution in various nucleon charges and pdf moments were found to be very small, and we expect the same here for the form factors. To check this we will compare results for various volumes with $M_{\pi}L$ values between 3 and 6.

Finally, we need to specify an upper bound on the pion momentum in the $N\pi$ state to truncate the sums in \pref{DefY} and \pref{DefZmu}. Following Refs.\ \cite{Bar:2016uoj,Bar:2016jof} we choose $|\vec{p}_n|\lesssim p_{\rm max}$ with $p_{\rm max}/\Lambda_{\chi}= 0.45$, where the chiral scale $\Lambda_{\chi}$ is equal to $4\pi f_{\pi}$. $N\pi$ states with pions satisfying this bound are called {\em low-momentum $N\pi$ states}. For these we expect our LO ChPT results to work reasonably well.\footnote{Recall that ChPT is an expansion in small pion momenta and masses.} States with pion momenta larger than the bound are called {\em high-momentum $N\pi$ states}. These too contribute to the excited-state contamination. However, choosing all euclidean time separations sufficiently large the contribution of the high-momentum $N\pi$ states is small and can be ignored.  The results in Refs.\ \cite{Bar:2016uoj,Bar:2016jof} suggest that at least a 1 fm separation between the operator and either source or sink is necessary. This corresponds to source-sink separations of 2 fm or larger in the 3-pt functions. We will take this time separation as a starting point to examine the range of applicability for our LO ChPT results. However, we will also argue that in case of the induced pseudoscalar form factor the results presented here can be applied at significantly smaller source-sink separations.

Note that an upper bound $|\vec{p}_n|\lesssim p_{\rm max}$ translates into a number $n_{p, {\rm max}}$ that depends on the spatial volume, i.e.\ on $M_{\pi}L$. The larger the volume the more discrete momenta satisfy the bound. Table \ref{table:npvalues} lists $n_{p, {\rm max}}$ for the volumes considered in this paper.\footnote{See also Ref.\ \cite{Bar:2017kxh} for the numbers corresponding to other values for the upper momentum bound.}

%%% Table 
\begin{table}[bt]
\begin{center}
\begin{tabular}{l|c|c|c|c|}
\multirow{2}{*}{$\frac{p_{{\rm max}}}{\Lambda_{\chi}}$}& \multicolumn{4}{c|}{$n_{p,{\rm max}}$ }  \\ 
& $M_{\pi}L=4$ & $M_{\pi}L=4$ & $M_{\pi}L=5$ & $M_{\pi}L=6$  \\  \hline
%0.3 & 1 & 2 & 4 & 5 \\
0.45& 3 & 5 & 8&12 
\end{tabular}
\caption{{\label{table:npvalues}} $n_{p,{\rm max}}$ and as a function of $p_{{\rm max}}/\Lambda_{\chi}$; see main text.}
\end{center}
\end{table}
%%%

\subsection{Impact on plateau estimates for the form factors}\label{ssect:impactonPlatest}

The effective form factors $G^{\rm eff}_{\rm A}(Q^2,t,t')$, $\GP^{\rm eff}(Q^2,t,t')$ in \pref{EffGA} and  \pref{EffGP} depend on the source-sink separation $t$ and the operator insertion time $t'$. For fixed $t$ we introduce the  plateau estimates that, as a function of $t'$, minimize the deviation from the true form factors. The results of the last section imply $\Delta G_{\rm A}(Q^2,t,t')>0$ and $\Delta \GP(Q^2,t,t')<0$, thus we define the plateau estimates according to
\begin{eqnarray}
G_{\rm A}^{\rm plat}(Q^2,t)&\equiv \min\limits_{0<t'<t}G_{\rm A}^{\rm eff}(Q^2,t,t')\,,\label{DefPlatEstGA}\\
\GP^{\rm plat}(Q^2,t)&\equiv \max\limits_{0<t'<t} \GP^{\rm eff}(Q^2,t,t')\,.\label{DefPlatEstGP}
\end{eqnarray}
These are functions of the momentum transfer and $t$. Naively one expects the operator has to be located closely to the middle between source an sink, i.e.\ $t'\approx t/2$.  At least for small momentum transfer that are accessible with ChPT we will find this expectation to be true. In practice, the midpoint estimates 
\begin{eqnarray}
G_{\rm A}^{\rm mid}(Q^2,t)&\equiv&G_{\rm A}^{\rm eff}(Q^2,t,t'=t/2)\,,\label{DefMidpointEstimateGA}\\
\GP^{\rm mid}(Q^2,t)&\equiv&\GP^{\rm eff}(Q^2,t,t'=t/2)\,.\label{DefMidpointEstimateGP}
\end{eqnarray}
are close to the plateau estimates and work equally well.

%==============
\begin{figure}[t]
\begin{center}
\includegraphics[scale=1.0]{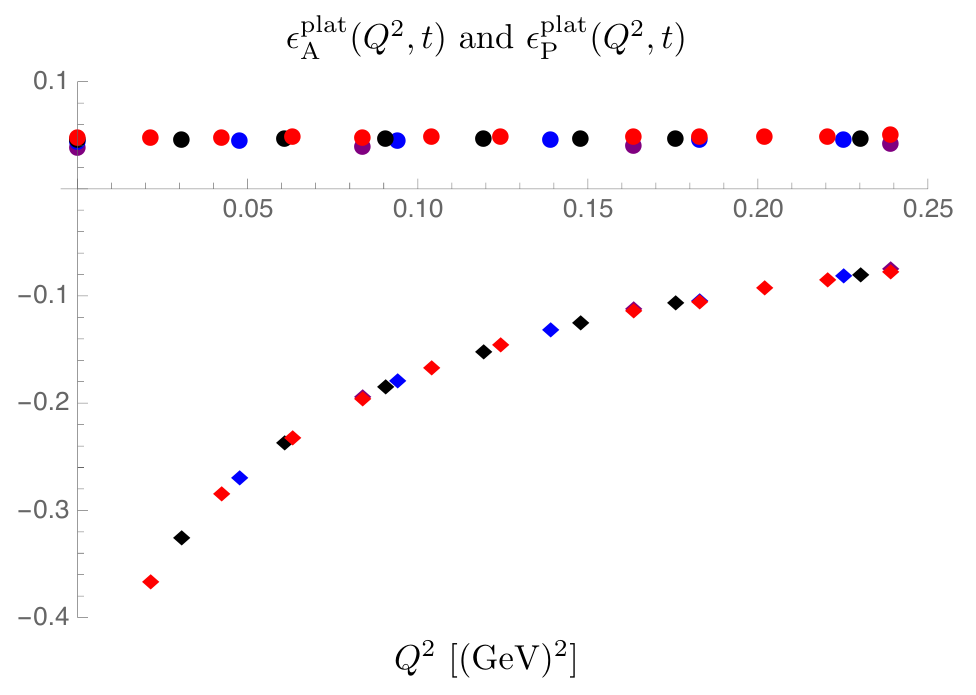}\\
\caption{\label{fig:epsilonAP2fm} Results for $ \epsilon^{\rm plat}_{\rm A}(Q^2,t)$ (dots) and $ \epsilon^{\rm plat}_{\rm P}(Q^2,t)$ (diamonds) for a source sink separation $t= 2$ fm and momentum transfers below $0.25\,\,{\rm (GeV)}^2$. The discrete values for the latter are determined by the size of the spatial volume given in terms of $M_{\pi}L=3$ (purple), 4 (blue), 5 (black) and 6 (red).}
\end{center}
\end{figure}
%============

As a measure for the $N\pi$-state contribution we introduce the relative deviation of the plateau estimates from the true form factors,
\begin{eqnarray}\label{DefEpsilons}
\epsilon^{\rm plat}_{\rm A}(Q^2,t)\equiv \frac{G^{\rm plat}_{\rm A}(Q^2,t)}{G_{\rm A}(Q^2)} -1\,,\quad \epsilon^{\rm plat}_{\rm {P}}(Q^2,t)\equiv \frac{\GP^{\rm plat}(Q^2,t)}{\GP(Q^2)} -1\,,
\end{eqnarray}
and analogously for the midpoint estimates. Figure \ref{fig:epsilonAP2fm} shows $\epsilon^{\rm plat}_{\rm A,P}$ for a source-sink separation of $t=2$ fm and small momentum transfers below $0.25\, {\rm GeV}^2$. Without the $N\pi$ contribution $ \epsilon^{\rm plat}_{\rm A,P}$ would be equal to 0. Any deviation from this value is the $N\pi$ state contamination in percent. 
Plotted are the results for the lowest discrete momentum transfers allowed by various spatial volumes with $M_{\pi}L$ values between 3 and 6. 

In case of the axial form factor (dots) we can read off that the plateau estimate  overestimates $G_{\rm A}(Q^2)$ by about 5\%, essentially independent of $Q^2$. We also reproduce the result for vanishing momentum transfer found in \cite{Bar:2016uoj}. 
In contrast, $\GP^{\rm plat}(Q^2)$ underestimates the induced pseudoscalar form factor by about 10\% to 40 \% (diamonds). The $Q^2$ dependence is rather pronounced, the smaller the momentum transfer the larger the deviation from the true form factor.

A small FV effect is noticeable in the data for $\epsilon^{\rm plat}_{\rm A}$. This is best seen by comparing the results for $M_{\pi}L=3$ and 6, which have some momentum transfers in common. The difference between these results is about half the size of the overlapping symbols. 
On the other hand, no FV effect is visible in $\epsilon^{\rm plat}_{\rm P}$. An explanation for this will be given below. 

Increasing the source-sink separation leads to a  smaller $N\pi$ contamination. For example, for $t= 3$ fm one roughly gains a factor 1/2: $\epsilon^{\rm plat}_{\rm A} $ drops to about $+2$\%, while $\epsilon^{\rm plat}_{\rm P}$ varies between  $-5$\% and $-20$ \%. The $Q^2$ dependence is qualitatively as in fig.\ \ref{fig:epsilonAP2fm}.

\begin{table}[tp]
\begin{center}
\begin{tabular}{|c||cc|cc||c|c|c|c||}
\hline
$n_q$ & $k_a$ & $\vec{n}_{q_a}$ &  $k_b$ & $\vec{n}_{q_b}$ & $X_{k_a}(\vec{q}_a,t,t'_{\rm ext})$ &  $X_{k_b}(\vec{q}_b,t,t'_{\rm ext})$ &\phantom{p} $ \epsilon_{\rm A}^{\rm plat}(Q^2,t)$ \phantom{p}& \phantom{p}$ \epsilon^{\rm plat}_{\rm P}(Q^2,t)$ \phantom{p} \\
\hline
\hgt 2 & 3 & (1,0,1) & 3 & (1,1,0)&0.207  & \phantom{$-$}0.045&0.045&$-$0.181\\
\hgt & 3 & (1,0,1) & 1 & (1,0,1)& 0.207 & $-$0.179 & 0.046 & $-$0.179 \\
\hgt  & 3 & (1,1,0) & 1 & (1,0,1)& 0.045 & $-$0.179 & 0.045 & $-$0.179\\
\hline
\hgt 5 & 3 & (2,0,1) & 3 & (2,1,0)& 0.077 &  \phantom{$-$}0.046 & 0.046 & $-$0.085\\
\hgt & 3 & (2,0,1) & 1 & (2,0,1)& 0.077 & $-$0.082 & 0.046 & $-$0.082 \\
\hgt  & 3 & (2,1,0) & 1 & (2,0,1)& 0.046 & $-$0.082 & 0.046 & $-$0.082\\\hline
\end{tabular}
\caption{{\label{table:nq23cases}} The $N\pi$ contributions $X_k(\vec{q},t,t'_{\rm ext})$ and $\epsilon^{\rm plat}_{\rm A,P}(Q^2,t)$ in the ratios and in the effective form factors, obtained from the three different ratio combinations specified in \pref{ratiocombi}. Results are shown for momenta with $n_q=2$ and $n_q=5$ for $M_{\pi}L=4$. The source-sink separation is $t=2$ fm in all cases, and $t'_{\rm ext}$ is between 0 and $t$ such that the $N\pi$-state contribution in the ratio is minimal.}
\end{center}
\end{table}

A few observations concerning the ChPT results are worth pointing out.
We already mentioned that the effective form factors for some momentum transfers can be obtained with different ratios and different 3-momenta $\vec{q}$. For instance, the two momenta $\vec{q}_A = \frac{2\pi}{L}(1,0,1)$ and $\vec{q}_B = \frac{2\pi}{L}(1,1,0)$ imply the same $Q^2$, and the effective form factors can be obtained from three inequivalent linear systems based on three combinations of ratios: 
\begin{eqnarray}
 R_{3}(\vec{q}_{A},t,t')  & {\rm and} & R_{3}(\vec{q}_{B},t,t')\,,\nn\\
 R_{1}(\vec{q}_{A},t,t')  & {\rm and} & R_{3}(\vec{q}_{B},t,t')\,, \label{ratiocombi}\\
 R_{1}(\vec{q}_{A},t,t') & {\rm and} & R_{3}(\vec{q}_{A},t,t')\,. \nn
\end{eqnarray}
It is not obvious that all three combinations lead to the same plateau estimates for the two form factors. 
One may expect one combination being afflicted with a smaller $N\pi$ contamination than the other two. However, it turns out that all three combinations give practically the same plateau estimates. Table \ref{table:nq23cases} summarizes the results for the example given in \pref{ratiocombi} for two momentum transfers. Apparently, the results for the plateau estimates, given in the last two columns, are essentially the same in all three cases. We conclude that in practice there is no reason to favor one case over the other.

Figure \ref{fig:epsilonAP2fm} shows that the $N\pi$ contribution in $\epsilon^{\rm plat}_{\rm A}$ is essentially independent of  the momentum transfer $Q^2$. Therefore, the tree diagrams that vanish identically for $Q^2=0$ seem to have no impact on the axial form factor for nonzero $Q^2$. In order to understand this we separate the total $N\pi$ contamination $X_{\mu}$ in eq.\ \pref{NpiConttot} according to their diagrammatic origin, i.e.\  we write
\begin{equation}\label{sepdiagcont}
X_{\mu}(\vec{q},t,t') = Z^{\rm tree}_{\mu}(\vec{q},t,t')+ Z^{\rm loop}_{\mu}(\vec{q},t,t') + \frac{1}{2}Y(\vec{q},t)\,.
\end{equation}
The first (tree) part contains the first two contributions in \pref{DefZmu} involving the coefficients $a_{\mu}$ and $\tilde{a}_{\mu}$, the second (loop) part the remaining three contributions in the same equation. The Y-part stems from the $N\pi$ contribution \pref{DefY} to the 2-pt function.

Table \ref{tableSepDiags} summarizes the relative deviations $\epsilon^{\rm mid}_{\rm A,P}$ for two momentum transfers based on a single origin only.\footnote{For the numbers in this table we have chosen the midpoint estimates \pref{DefMidpointEstimateGA}, \pref{DefMidpointEstimateGP}, for simplicity. For the conclusions drawn in this section this simplification is irrelevant.}
As expected the 2-pt function contribution is small, much smaller than the contributions from the 3-pt function. In contrast to the latter there are no exponentials involving the shorter time separations $t'$ and $t-t'$ in the 2-pt function, thus the $N\pi$ contribution in the 2-pt function is exponentially more suppressed.

\begin{table}[tp]
\begin{center}
\begin{tabular}{|c||cc|cc||c|c|c||c|c|c||}
\hline
$n_q$ & $k_a$ & $\vec{n}_{q_a}$ &  $k_b$ & $\vec{n}_{q_b}$ & $ \epsilon^{\rm mid}_{\rm A, tree}(Q^2_n,t)$ & $\epsilon^{\rm mid}_{\rm A, loop}(Q^2_n,t)$ &$\epsilon^{\rm mid}_{\rm A, 2pt}(Q^2_n,t)$ &$ \epsilon^{\rm mid}_{\rm P, tree}(Q^2_n,t)$ & $\epsilon^{\rm mid}_{\rm P, loop}(Q^2_n,t)$ &$\epsilon^{\rm mid}_{\rm P, 2pt}(Q^2_n,t)$\\
\hline
\hgt 2 & 3 & (1,0,1) & 3 & (1,1,0)&0.000&  0.048& $-$0.003 & $-$0.179& $-$0.002& $-$0.003\\
\hgt & 3 & (1,0,1) & 1 & (1,0,1)& 0.000 & 0.049& $-$0.003  & $-$0.179& $-$0.000& $-$0.003 \\
\hgt  & 3 & (1,1,0) & 1 & (1,0,1)&0.000 &0.048 & $-$0.009 &$-$0.179& $-$0.000&$-$0.003\\
\hline
\hgt 5 & 3 & (2,0,1) & 3 & (2,1,0)& 0.000& 0.050& $-$0.008& $-$0.087& $-$0.004& $-$0.008\\
\hgt & 3 & (2,0,1) & 1 & (2,0,1)& 0.000 & 0.050& $-$0.008& $-$0.087& $-$0.000& $-$0.008\\
\hgt  & 3 & (2,1,0) & 1 & (2,0,1)& 0.000& 0.050& $-$0.008 & $-$0.087& $-$0.000& $-$0.008\\
\hline\hline
\end{tabular}
\caption{{\label{tableSepDiags}}The relative deviations $\epsilon^{\rm mid}_{\rm X, tree}, \epsilon^{\rm mid}_{\rm X,loop}$ and $\epsilon^{\rm mid}_{\rm X, 2pt}$ for two different momentum transfers. The source-sink separation is set to $t=2$ fm and and $M_{\pi}L=4$. }
\end{center}
\end{table}

More striking is the following observation. In the axial form factor the tree diagrams do not contribute, the entire $N\pi$ contamination stems from the loop diagrams. For the induced pseudoscalar form factor it is the other way around, the loop diagrams do not contribute, the dominant contribution has its origin in the tree diagrams.
 
To understand this consider the case where the momenta and ratios are chosen in such a way that the matrix $M$ in \pref{DefMmatrix}  is diagonal. Two explicit examples are given in the third and sixth row of table \ref{table:nq23cases}.  In these cases  $G_{\rm A}^{\rm mid} $  is proportional to $R_3$ with a 3-momentum having $q_{3}=0$. For such a momentum the coefficients $a_{3}(\vec{q})$ and $\tilde{a}_{3}(\vec{q})$ vanish according to eqs.\ \pref{DefaAndat} and \pref{resLimaAndat} -- \pref{atildecorrk}. Thus, the tree contribution $\epsilon^{\rm mid}_{\rm A,{\rm tree}}$ vanishes identically. The loop contribution is dominated by the contribution proportional to the coefficients $B_3^{\infty}$ and $\tilde{B}_3^{\infty}$ in \pref{resLimB3} and \pref{resLimBtk}. For spatial momenta with $q_3=0$ the $\vec{q}$ dependence cancels exactly and the coefficients are essentially the ones for $\vec{q}=0$. Therefore, the $N\pi$ contamination in $G_{\rm A}^{\rm mid}(Q^2,t) $ is essentially as in the axial charge, the form factor for $Q^2=0$.
 
On the other hand, $\GP^{\rm mid}$ is proportional to $R_1$. Therefore, the dominant tree contribution stems from $a_1^{\infty}=\tilde{a}_1^{\infty}=-1/2$, c.f.\ \pref{resLimaAndat}. This value is rather large and negative, explaining the underestimation of $\GP(Q^2)$ displayed in fig.\ \ref{fig:epsilonAP2fm}. The loop contribution is governed by the coefficients $B_1^{\infty}$ and $\tilde{B}_1^{\infty}$ in \pref{resLimBk} and \pref{resLimBtk}. Except for the opposite sign these coefficients are of the same size as $B_3^{\infty}$ and $\tilde{B}_3^{\infty}$, which are responsible for the non-vanishing loop contribution $\epsilon^{\rm mid}_{\rm A,loop}$. The key observation is that even though the individual contribution for one particular pion momentum $\vec{p}$ is non-zero, the sum over all momenta $\vec{p}$ with the same $|\vec{p}| $ vanishes because the coefficient $B_1^{\infty}$ is proportional to $p_{1}p_{3}$, 
\begin{equation}\label{sumB1coeff}
\sum_{\vec{p},|\vec{p}|={\rm fix}} B_1^{\infty}(\vec{q},\vec{p})=0\,.
\end{equation}
In the $N\pi$ contribution to the ratios the sum also involves the exponentials with the energy gaps, and including these in \pref{sumB1coeff} leads to a nonzero but small number on the right hand side. In contrast, the coefficient $B_3^{\infty}$ is proportional to $p^2 +p_{3}^2$, thus performing the sum \pref{sumB1coeff} with this coefficient we sum up positive numbers only and end up with a sizeable nonzero result. 

The particular results for the coefficients ``explain'' why the tree diagrams do not contribute to $\epsilon_{\rm A}$ and the loop diagrams not to $\epsilon_{\rm P}$. However, this explanation becomes less transparent when $\epsilon_{\rm P}$ is obtained only from the ratios $R_3$ with two different spatial momenta, as in the first and fourth rows in table \ref{table:nq23cases}. In these cases one of the ratios $R_3$ is evaluated with a momentum with $q_3\neq 0$. Thus, there is a non-vanishing positive tree contribution caused by $a_3(\vec{q})$ and $\tilde{a}_3(\vec{q})$ leading to a larger $N\pi$ contribution than in the ratio with a momentum satisfying $q_3=0$ (see first and fourth row in table \ref{table:nq23cases}). Nevertheless, the linear system that needs to be solved to obtain $G^{\rm eff}_{\rm P}$ is such that the  overestimation in the two ratios $R_3$  result in an underestimation in  $G^{\rm eff}_{\rm P}$ that equals the one obtained  in the direct determination based on an a combination of ratios involving $R_1$.

%==============
\begin{figure}[t]
\begin{center}
\includegraphics[scale=1.0]{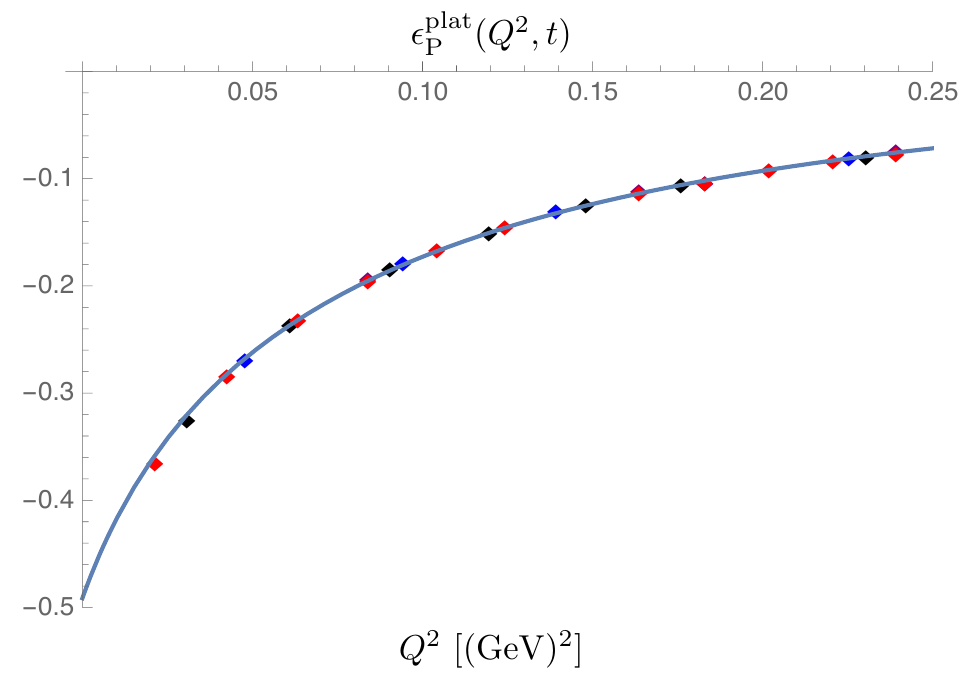}\\
\caption{\label{fig:epsilonAP2fmwApprox} Results for $ \epsilon^{\rm plat}_{\rm P}(Q^2,t)$ as in fig.\ \ref{fig:epsilonAP2fm}. The solid line corresponds to the approximation in \pref{epsGPapp}.}
\end{center}
\end{figure}
%============

That only the tree diagrams contribute to the $N\pi$ contamination in $\epsilon^{\rm plat}_{\rm P}(Q^2,t)$ has two consequences. Firstly, it is independent of the spatial volume, because the tree diagram contribution does not explicitly depend on the spatial extent $L$, see eqs.\ \pref{DefaAndat} to \pref{at4}. Consequently, the results for different $M_{\pi}L$ values in fig.\ \ref{fig:epsilonAP2fm} fall on an smooth curve. Secondly, the dominant contribution stems from diagram c) in fig.\ \ref{fig:diagsSN} only, the contribution from diagrams m) and n) in fig.\ \ref{fig:Npidiags} is O($1/M_N$) suppressed. Keeping only the dominant contribution an excellent approximation for $\epsilon^{\rm plat}_{\rm P}(Q^2,t)$ is given by the simple expression
\begin{equation}\label{epsGPapp}
\epsilon^{\rm plat}_{\rm P}(Q^2,t)\, \approx\, -\exp\left(-E_{\pi,\vec{q}}\frac{t}{2}\right) \,.%\cosh\left[ \frac{\vec{q}^{\,2}}{2 M_N} \frac{t}{2}\right]\,.
\end{equation}
Figure \ref{fig:epsilonAP2fmwApprox} shows again the results of figure \ref{fig:epsilonAP2fm} together with the approximation \pref{epsGPapp}. Obviously, the simple expression captures $\epsilon^{\rm plat}_{\rm P}(Q^2,t)$ very well. 
The right hand side of \pref{epsGPapp} depends only on the source-sink separation and the energy of a pion with spatial momentum $\vec{q}$. The maximal deviation $-\exp[-M_{\pi}t/2]$ is assumed in the limit of vanishing momentum transfer.

It is conceivable that the ChPT result for  $\epsilon^{\rm plat}_{\rm P}(Q^2,t)$ can be applied at source-sink separations substantially smaller than 2 fm. Recall that this bound was imposed to guarantee a sufficient exponential suppression of the high-momentum pion states in the loop diagrams. Since the loop contribution in $\epsilon^{\rm plat}_{\rm P}(Q^2,t)$ essentially cancels it seems plausible that the bound for a minimal source-sink separation can be relaxed significantly. We will study this issue in section \ref{sec:confr} when we compare the ChPT results with actual lattice data. 

%========================
\subsection{Impact on the ratio $R_4$}\label{sect:ImpactA4}
%========================

We have already mentioned that data for the ratio $R_4(\vec{q},t,t')$ is usually not taken into account in lattice calculations of the nucleon form factors \cite{Capitani:2017qpc,Alexandrou:2017hac}. On one hand the 3-pt function involving the time component $A_4$ of the axial vector current is found to be statistically noisy, noisier than for the spatial components. In addition, the excited-state contamination in $R_4(\vec{q},t,t')$ is found to be much more severe than in the other ratios. 

\begin{figure}[t]
\begin{center}
\includegraphics[scale=1.0]{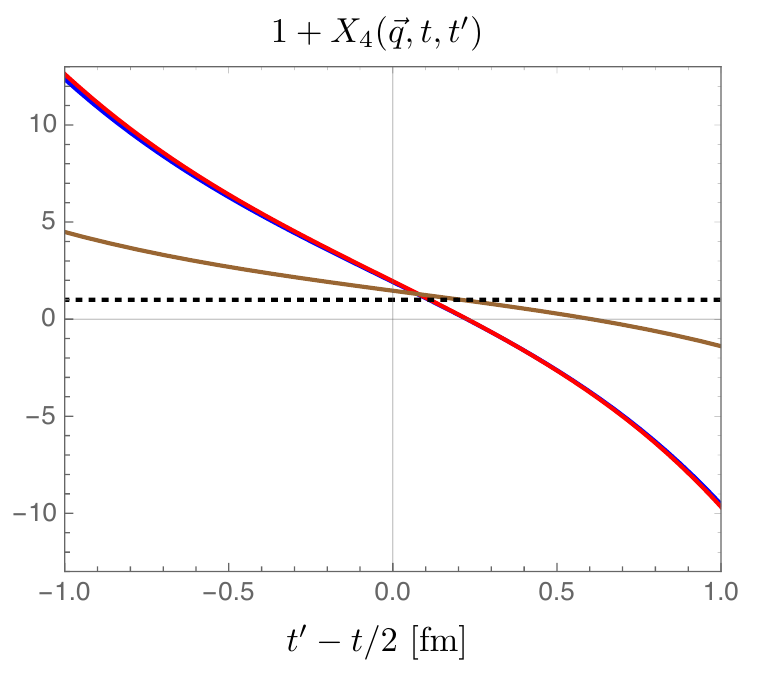}\\
\caption{\label{fig:R4ChPT2fm} The  factor $1+X_{4}(\vec{q},t,t')$ (red) entering the ratio $R_4(\vec{q},t,t')$ in \pref{NpiConttot} as a function of the shifted operator insertion time $t'-t/2$. Result for $t=2$ fm, a spatial volume $M_{\pi}L=4$ and the smallest non-vanishing momentum with $n_{\vec{q}}=(0,0,1)$. Also shown the approximation $1+Z^{\rm tree}_{4}(\vec{q},t,t')$ (blue) for the same parameters. Analogous result (brown) for twice the source-sink separation, $t=4$ fm. The black dotted line shows the expected constant result for $t\rightarrow \infty$.}
\end{center}
\end{figure}

Figure \ref{fig:R4ChPT2fm} shows $1+X_4(\vec{q},t,t')$ (red line) for a fixed source-sink separation $t=2$ fm as a function of the (shifted) operator insertion time $t'-t/2$. The spatial volume is such that $M_{\pi}L=4$ and a momentum transfer $\vec{q}$ with $n_q=(0,0,1)$. Recall that $X_4(\vec{q},t,t')$
vanishes for both time separations taken to infinity, implying that for asymptotically large $t$ fig.\ \ref{fig:R4ChPT2fm} should show a horizontal line at +1 (black dotted line). Instead we observe an almost linear dependence on $t'$ with a sizeable negative slope. Note that this behavior does not allow to define a plateau estimate for $R_4(\vec{q},t,t')$ as we have done for the effective form factors in \pref{DefPlatEstGA} and \pref{DefPlatEstGP}. 

The first observation we can make is that the dominant $N\pi$ contribution stems from the tree diagrams, i.e.\ $X_4(\vec{q},t,t')\approx Z^{\rm tree}_4(\vec{q},t,t')$. The latter is displayed by the blue line in fig.\ \ref{fig:R4ChPT2fm}. The loop contributions average away when the sum over the pion momenta is taken, the argument is the same as the one given for the form factor $\GP(Q^2,t,t')$ in the previous section, c.f.\  \pref{sumB1coeff}. The contribution $Y(\vec{q},t)$ from the 2-pt function is also small for $t=2$ fm.

Since the loop diagrams do not contribute to the $N\pi$ contribution in $R_4(\vec{q},t,t')$ we expect to be able to relax our bound $t \gtrsim 2$ fm that we imposed to suppress the high-momentum $N\pi$ states. Only one $N\pi$ state with a small pion momentum $|\vec{q}|$ contributes and the ChPT result is expected to be applicable for source-sink separations smaller than 2 fm. However, recall that ChPT is not expected to work when the operator is close to either source or sink. In other words, in fig.\ \ref{fig:R4ChPT2fm} we should focus on the region with $|t'-t/2| \approx 0$. 

The qualitative behavior seen in fig.\ \ref{fig:R4ChPT2fm} is easily understood. It can be traced back to the relative sign between the coefficients $a^{\infty}_{4}(\vec{q})$ and $\tilde{a}^{\infty}_{4}(\vec{q})$ given in \pref{a4} and \pref{at4}. Taking only these leading coefficients into account in $Z^{\rm tree}_4(\vec{q},t,t')$ and dropping the small energy difference $E_{N,\vec{q}} - M_N$ we approximately find
\begin{equation}\label{Z4treeApprox}
Z^{\rm tree}_4(\vec{q},t,t') \approx -\frac{2M_NE_{\pi,\vec{q}}}{M_{\pi}^2} \exp\left(-\frac{E_{\pi,\vec{q}}\, t}{2}\right) \sinh\left(E_{\pi,\vec{q}}\left(t'-\frac{t}{2}\right)\right)\,,
\end{equation}
and it is essentially this $-\sinh\left(E_{\pi,\vec{q}}\left(t'-\frac{t}{2}\right)\right)$ behavior we observe in fig.\ \ref{fig:R4ChPT2fm}. 

Note that the prefactor \pref{Z4treeApprox} is numerically fairly large, mainly due to the factor $M_N/M_{\pi}$. The SN contribution in $R_4(\vec{q},t,t')$ is O($1/M_N$) suppressed compared to the $N\pi$ contribution, thus much larger source-sink separations are necessary to suppress these. Fig.\ \ref{fig:R4ChPT2fm} also shows the result for twice the source-sink separation $t=4$ fm (brown line). Even at this large time separation a non-negligible slope is still visible.
To make this more quantitative let us introduce the mid-point estimate
\begin{equation}\label{midpointR4}
\Pi^{\rm mid}_4(\vec{q},t) = R_4(\vec{q},t,t'=t/2) = \Pi_4(\vec{q})[1+X_4(t,t'=t/2)]
\end{equation}
for the constant $\Pi_4(\vec{q})$ we are interested in. The sizeable $N\pi$ contributions manifests in poor mid-point estimates. At $t=2$ fm \pref{midpointR4} overestimates by about 90\%, and this number decreases to about 45\% for $t=4$ fm. 

%========================
\section{Comparison with lattice data}\label{sec:confr} 
%========================

\subsection{Preliminaries}
To compare the ChPT results of this paper with lattice QCD data we ideally need continuum extrapolated data with a (near to) physical pion mass. The spatial volume should be sufficiently large with $M_{\pi}L\simeq 3$ or larger and the data must have been obtained with the plateau method.\footnote{Many lattice collaborations resort to the summation method \cite{Maiani:1987by,Capitani:2012gj} or employ multi-exponential fits in their data analysis to suppress or explicitly account for the excited-state contamination. The ChPT results presented here cannot be applied to such data.} Finally, the time separations in the correlation functions need to be sufficiently large such that they are dominated by pion physics.

The last issue is the real bottleneck for a comparison with lattice data. Source-sink separations of 2 fm and larger, as we require for the axial from factor $G_{\rm A}$, are out of reach with current simulation techniques. However, we argued that our ChPT might be applicable at significantly smaller time separations in case of $\GP$ and the ratio $R_4$. For that reason we focus on these two quantities in the following and compare with recently published data that roughly match our requirements \cite{Ishikawa:2018rew,Bali:2018qus}.

\subsection{Induced pseudoscalar form factor}\label{Ipsff}

In Ref.\ \cite{Ishikawa:2018rew} the PACS collaboration reports plateau estimate data for the two nucleon form factors. 
The results were obtained on a $96^4$ lattice with lattice spacing $a\approx 0.085$ fm and a pion mass $M_{\pi}\approx146$ MeV. The spatial lattice extent $L\approx 8.1$ fm is rather large, corresponding to $M_{\pi}L\approx 6.0$. The source-sink separation equals 15 time slices,  i.e.\ $t\approx1.3$ fm, and the central four time slices with $6\le t/a \le 9$ were used to obtain the plateau estimates. For more simulation details see \cite{Ishikawa:2018rew}.
 
\begin{figure}[t]
\begin{center}
\includegraphics[scale=1.0]{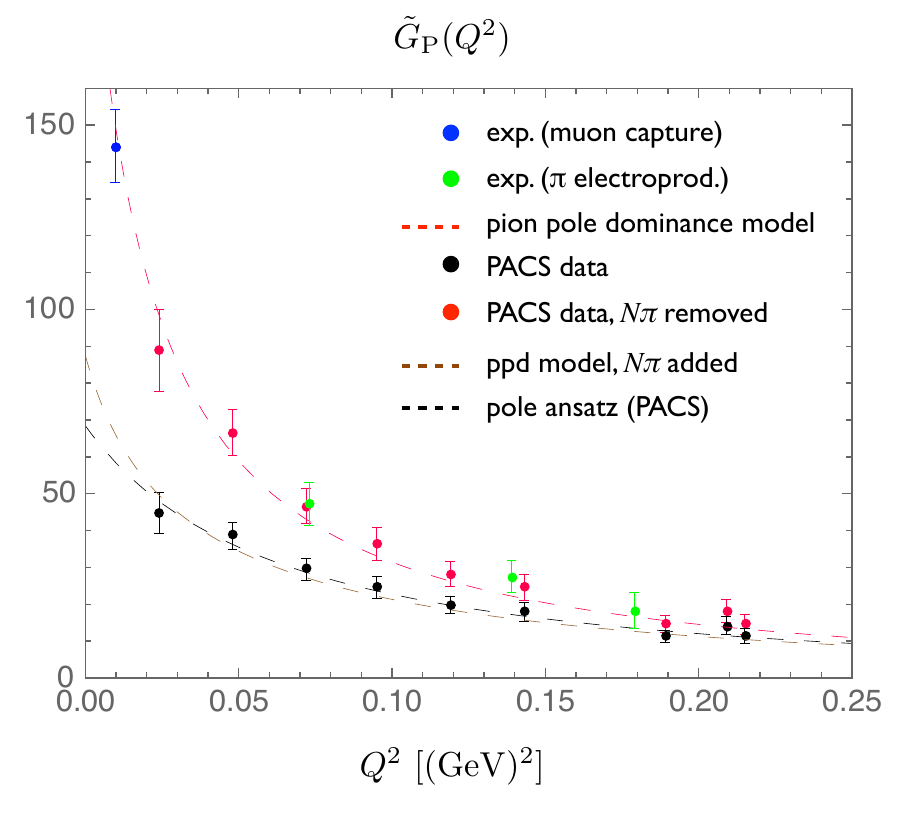}\\
\caption{PACS data for the momentum transfer dependence of the renormalized induced pseudoscalar form factor.  Black symbols are the original plateau estimate data given in Ref.\ \cite{Ishikawa:2018rew}.  Red symbols correspond to the data corrected according to eq.\ \pref{pacsdatacorrected}. The prediction of the ppd model is given by the red dashed line, while the corrected ppd model is shown by the dashed brown line. The pole-ansatz description according to \pref{poleAnsatz} of the original PACS data is given by the black dashed line. }
\label{fig:pacsplotcorr2}
\end{center}
\end{figure}

Figure \ref{fig:pacsplotcorr2} shows essentially fig.\ 16 of Ref.\ \cite{Ishikawa:2018rew}. It displays the numerical PACS results for the renormalized induced pseudoscalar form factor (black data points) together with existing experimental results (blue and green data points) and the analytic expectation by the pion-pole-dominance (ppd) model (red dashed line). In this model the two form factors are given by
\begin{equation}
\GP(Q^2) \approx \frac{4M_N^2 G_{\rm A}(Q^2)}{Q^2 + M_{\pi}^2}\,,\qquad G_{\rm A}(Q^2) \approxÊ\frac{G_{\rm A}(0)}{(1+Q^2/{M}_{\rm A}^2)^2} \,.
\end{equation}
In Ref. \cite{Ishikawa:2018rew} the value ${M}_{\rm A}^2\approx1.04$ GeV was chosen, stemming from $ r_{\rm A}^2 = 12/ {M}_{\rm A}^2$ with $r_{\rm A}\approx0.67$ fm.  
 
%=========
% Result 6
%=========
%========= 

For small momentum transfers the lattice data are incompatible with the ppd model and the experimental data. The PACS collaboration found that the data are well described by a ppd-inspired ansatz (black dashed line in fig.\ \ref{fig:pacsplotcorr2})
\begin{equation}\label{poleAnsatz}
\GP(Q^2) \approx \frac{4M_N^2 G_{\rm A}(Q^2)}{Q^2 + M_{\rm pole}^2}\,,
\end{equation}
with $M_{\rm pole}=256(17)$ MeV determined by a fit of \pref{poleAnsatz} to the data. This mass is about twice as large as the pion mass in the simulation.

The plateau estimates were obtained at a single source-sink separation $t\approx 1.3$ fm. For such a small time separation we can expect the plateau estimates to differ significantly from the physical values  at $t=\infty$ due to the presence of excited states. With our result $\epsilon^{\rm plat}_{\rm P}(Q^2,t)$  we can correct the data and analytically remove the anticipated  LO $N\pi$-state contamination by calculating 
\begin{equation}\label{pacsdatacorrected}
\GP^{\rm corr}(Q^2,t) \equiv \frac{\GP^{\rm plat}(Q^2,t) }{1+ \epsilon^{\rm plat}_{\rm P}(Q^2,t)},
\end{equation}
setting $t=1.3$ fm. Provided higher order corrections and other excited-state contributions are small we expect
\begin{equation}\label{tindepofGPcorr}
\GP^{\rm corr}(Q^2,t) \approx \GP(Q^2)\,,
\end{equation}
i.e.\ the corrected data should be close to the true form factor.
 
To correct the data we can use the simple approximation in \pref{epsGPapp}, and the result is shown in fig.\ \ref{fig:pacsplotcorr2} by the red symbols. The corrected lattice data are in good agreement with the experimental data and the ppd model. In fact, the improvement is better than naively expected. For source-sink separations as small as 1.3 fm one would not be surprised if excited states other than two-particle $N\pi$ states also contribute and distort the form factor. That the correction works very well at $t =1.3$ fm supports our expectation that the ChPT results for $G^{\rm plat}_{\rm P}(Q^2,t)$ are applicable for source-sink separations well below 2 fm. 

Instead of correcting the lattice data by removing the $N\pi$-state contamination we can also correct the ppd model for the presence of the $N\pi$-state contamination, 
 \begin{equation}\label{ppdcorrection}
G^{\rm ppd+N\pi}_{\rm P}(Q^2,t) = G^{\rm ppd}_{\rm P}(Q^2) \big[ 1+  \epsilon^{\rm plat}_{\rm P}(Q^2,t)\big]\,.
\end{equation}
The resulting curve $G^{\rm ppd+N\pi}_{\rm P}(Q^2,t) $ is also shown in fig.\ \ref{fig:pacsplotcorr2} (brown dashed line).  It is nearly indistinguishable from the pole ansatz result \pref{poleAnsatz} found by the PACS collaboration to describe the data very well.

The correction formula \pref{pacsdatacorrected} and its region of applicability needs to be carefully studied before it can be applied to extract the physical form factor $\GP(Q^2)$ from lattice data. For this data at various source-sink separations will be extremely useful, since these will allow to check whether the corrected data are indeed independent of the source-sink-separation, i.e.\ whether eq.\ \pref{tindepofGPcorr} is satisfied.

%=========
\begin{figure}[p]
\begin{center}
\includegraphics[scale=1.0]{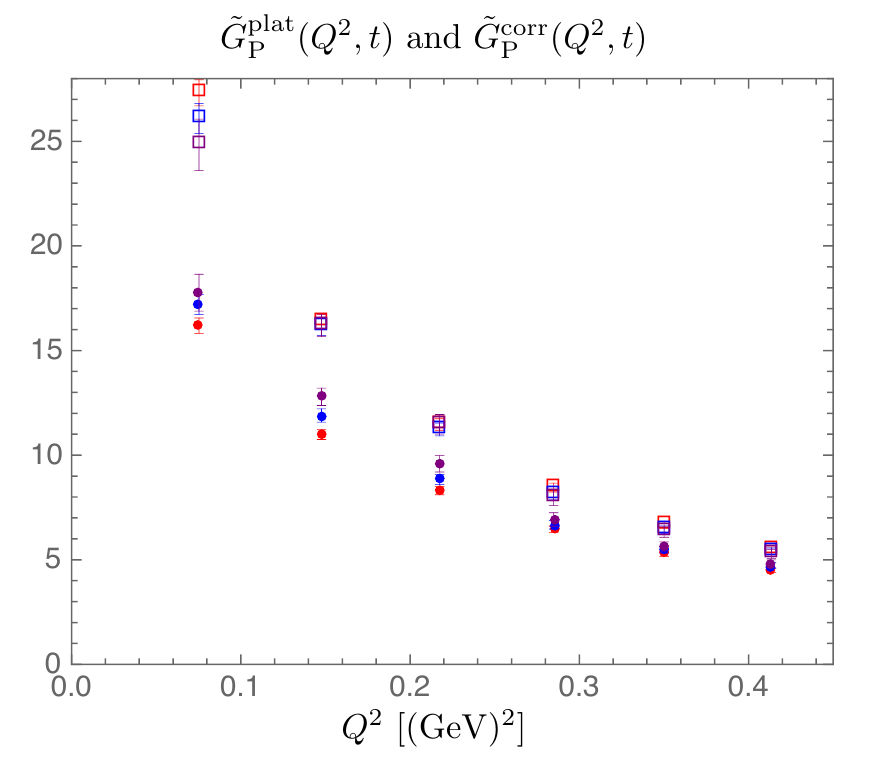}\\
\caption{Plateau estimate data (solid circles) of Ref.\ \cite{Alexandrou:2017hac} and corrected data (open squares) for the renormalized induced pseudoscalar form factor for three source-sink separations: $t=0.94$ (red), $t=1.13$ fm (blue) and $t=1.31$ fm (purple).  }
\label{fig:etmcGPofQ2}
\end{center}
\end{figure}
%=========
\begin{figure}[p]
\begin{center}
\includegraphics[scale=1.0]{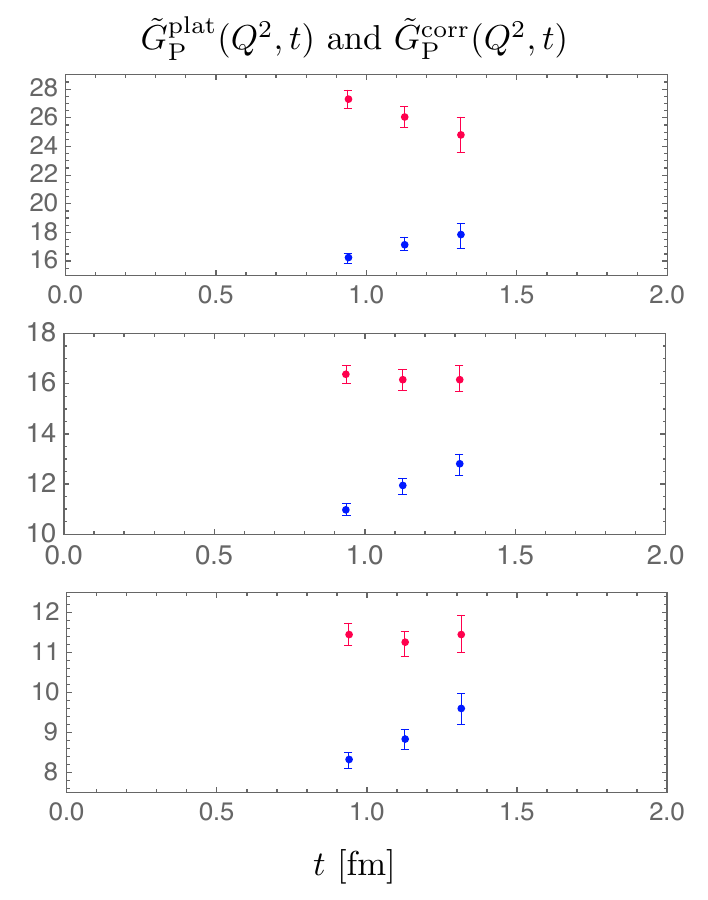}\\
\caption{Data of fig.\ \ref{fig:etmcGPofQ2} as a function of the source-sink separation $t$ for the smallest three momentum transfers: $Q^2=0.073$ GeV$^2$ (upper panel), 0.148 GeV$^2$ (middle panel) and 0.217 GeV$^2$ (lower panel). Original plateau data in blue, corrected data in red.  }
\label{fig:etmcGPofts}
\end{center}
\end{figure}
%========= 
%========= 

Ref.\ \cite{Alexandrou:2017hac} reports plateau estimate data for three source-sink separations $t=0.94, 1.13$ and $1.31$ fm. The lattice ensemble was generated with two-flavor twisted mass fermions with a pion mass $M_{\pi}\approx130$ MeV and a lattice spacing $a\approx 0.094$ fm. The spatial volume is somewhat small satisfying $M_{\pi}L=2.98$. For more details about the simulation parameters we refer to \cite{Alexandrou:2017hac}.

Figure \ref{fig:etmcGPofQ2} shows the plateau estimate data (solid circles) and the corrected data (open squares) for the lowest six momentum transfers for all three source-sink separations.\footnote{I thank C.\ Alexandrou for sending me the plateau estimate data.} For the lowest three $Q^2$ values the original data exhibit a clear dependence on the source-sink separation, while the corrected data are compatible with being constant as a function of $t$. This is better seen in fig. \ref{fig:etmcGPofts} where the data are shown as a function of $t$. The corrected data (red symbols) in the lower two panels are compatible with being $t$ independent, while the original plateau data (blue symbols) show a clear trend to increase as $t$ becomes larger.  For the smallest $Q^2$ value in the upper panel this is not as convincing as for the next two larger $Q^2$ values, but the statistical error is also larger in this case. 

The range covered by the three source-sink separations in fig.\ \ref{fig:etmcGPofts} is rather small, and more data for larger $t$ values would be beneficial to test the correction formula. Larger source-sink separations are hard to achieve in simulations with a (near to) physical pion mass. Ref.\ \cite{Green:2017keo} reports data obtained from a single 2+1 flavor ensemble with clover-improved Wilson fermions. The pion mass 317 MeV is rather heavy and the lattice spacing $a\approx 0.11$ fm is also rather course. The volume, however, is quite large satisfying $M_{\pi}L\approx 5.9$. For more details see Ref. \cite{Green:2017keo}. 

%=========
% Result 7
%=========
\begin{figure}[bt]
\begin{center}
\includegraphics[scale=0.9]{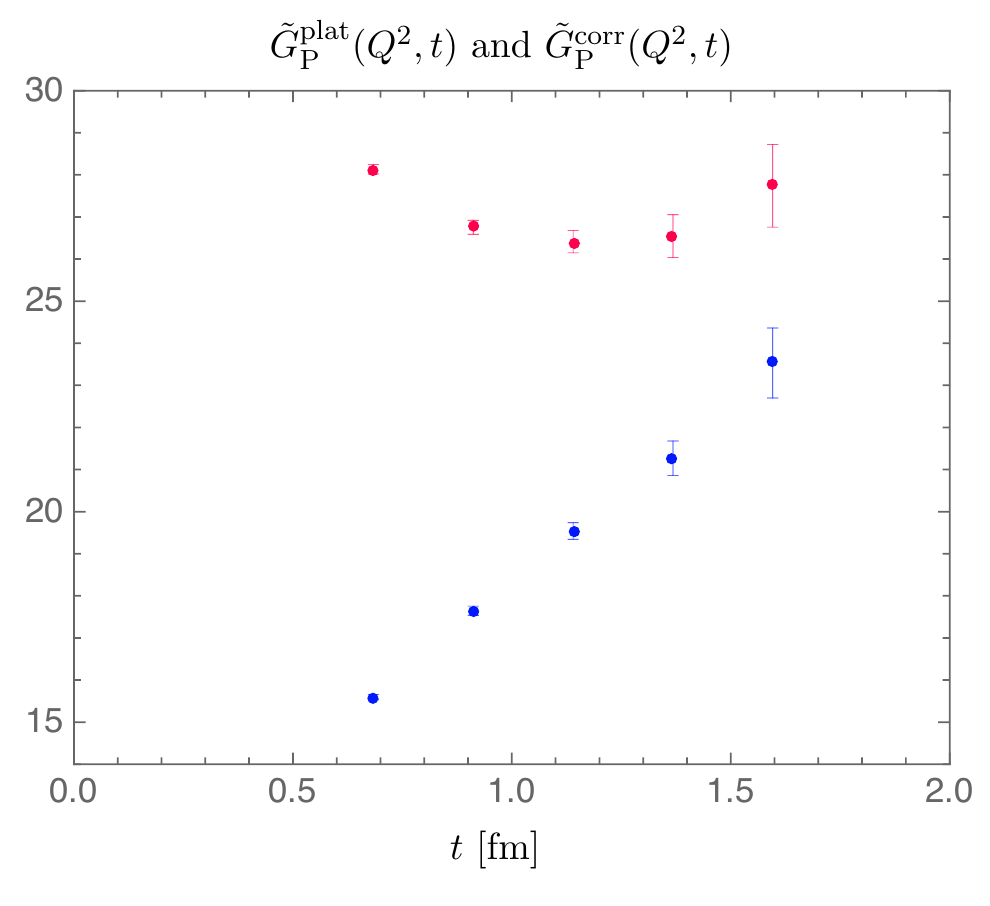}\\
\caption{Plateau estimate data (blue symbols) of Ref.\ \cite{Green:2017keo} and corrected data (red symbols) as a function of the source-sink separation $t$. }
\label{fig:plotGreen}
\end{center}
\end{figure}
%========= 

Figure \ref{fig:plotGreen} shows the plateau estimate data $\GP^{\rm plat}(Q^2,t)$ (blue symbols) for one momentum transfer $Q^2=0.12$ GeV$^2$, as it is displayed in Fig.\ 5 of Ref.\ \cite{Green:2017keo}.\footnote{I thank J.\ Green for sending me the data.} To a good approximation the data show a linear rise as a function of $t$. The corrected data (red symbols), on the other hand, seem to reach a plateau at $t$ around 1 fm, but the data point at the largest source-sink separation is slightly too large. Still, taking into account that the pion mass is rather heavy the simple correction formula works surprisingly well. 

Our comparison between lattice plateau estimate data and the ChPT results for the $N\pi$ state contamination in it is not conclusive. More data at larger source-sink separations are needed  to corroborate the ChPT results presented here, in particular to validate the correction formula \pref{pacsdatacorrected} for the removal of the $N\pi$-state contamination from lattice data. Recently, the PACS collaboration published form factor data obtained from an ensemble with a 135 MeV pion mass and a finite volume of size $(10.8 {\rm \, fm})^4$. Plateau estimates for the form factors exist for four source-sink separations, with the largest one of about 1.35 fm. A dependence on the source-sink separation is clearly visible and it is certainly interesting to compare the data with the ChPT predictions presented here.\footnote{Unfortunately, the data are not publicly available yet \cite{EigoShintaniPC}.} Nevertheless, the main conclusion we can draw so far is that the $N\pi$-state contamination in $G^{\rm plat}_{\rm P}(Q^2)$ results in a softening of the anticipated ppd behavior at small $Q^2$, a feature that has been observed in many lattice results so far.

%========================
\subsection{The ratio ${\mathbf \it R_4(Q^2, t,t')}$}
%========================

In a recent paper \cite{Bali:2018qus} RQCD presented data for the ratio $R_4(Q^2, t,t')$ involving the time component $A_4$ of the axial vector current.\footnote{Ref.\ \cite{Bali:2018qus} is mainly formulated assuming the Minkowski metric, thus the subscript 0 is used for the time component.}
It was observed that the excited-state contribution is strongly enhanced compared to the SN ground-state contribution. The enhancement was so strong that a standard multi-state fit ansatz failed to account for the excited-state contribution. 

In section \ref{sect:ImpactA4} we discussed the peculiar features of the ratio $R_4$ compared to its spatial counterparts. In particular, we emphasized that the SN contribution is $1/M_N$ suppressed compared to the $N\pi$ contribution. Therefore it is interesting to check whether the excited-state effects in the data of Ref.\ \cite{Bali:2018qus} can be attributed to $N\pi$ states.

RQCD analyzed data obtained with two-flavor non-perturbatively improved Wilson fermions. The ensemble with the lightest pion has $M_{\pi}\approx 150$ MeV, a lattice spacing $a\approx 0.071$ fm and a finite volume satisfying $M_{\pi}L\approx 3.47$. Data for the ratios are available for three source-sink separations corresponding to $t=9a, 12a$ and $15a$. In the following we focus on the largest value $t\approx 1.07$ fm, which, for our purposes, is still rather small.    

\begin{figure}[t]
\begin{center}
\includegraphics[scale=1.0]{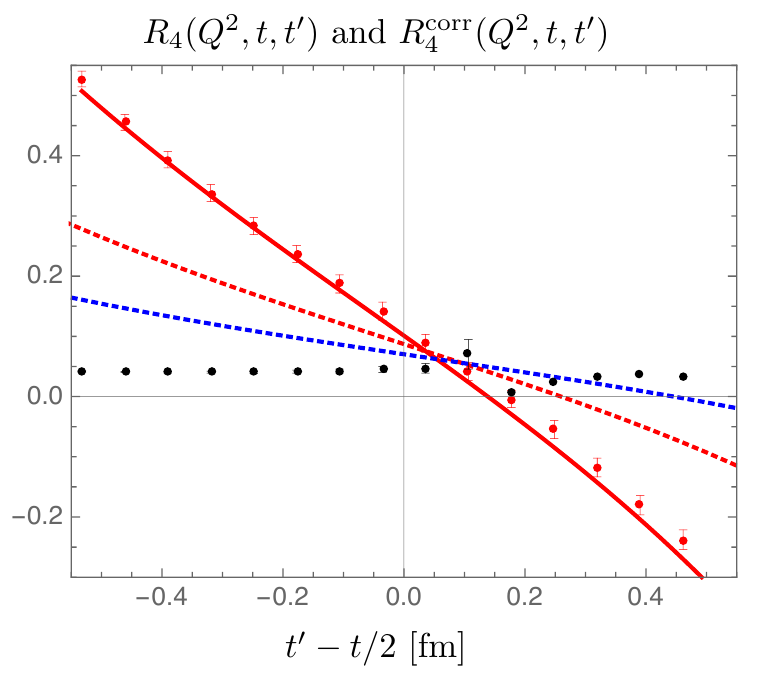}\\
\caption{\label{fig:2} Data for the correlation function ratio $R_4(Q^2,t,t')$ (red data points) and corrected data (black symbols) according to \pref{R4corrected} for $t\approx 1.07$ fm and $Q^2=0.073$ GeV$^2$. The ChPT result is given by the red solid line. The dotted lines are the ChPT results for $t\approx 2.1$ fm (red) and $t\approx 3.2$ fm (blue).}
\end{center}
\end{figure}

Figure \ref{fig:2} shows the data for $R_4(Q^2, t,t')$ (red data points) as a function of the (shifted) operator insertion time $t'-t/2$ for fixed $t=1.07$ fm and for fixed momentum transfer $Q^2 =0.073$ GeV$^2$.\footnote{I thank T.~Wurm for sending me the data. In Ref.\ \cite{Bali:2018qus} the data are displayed in figure 6, left panel.} The data do not exhibit a plateau and show roughly a linear dependence on $t'$.\footnote{A similarly looking plot was shown by T.~Schulz at the conference {\em Lattice 2018} \cite{TschulzLat18}.} 

The LO ChPT result for the ratio is shown by the red solid line in fig.\ \ref{fig:2}. The line describes the data very well. Recall that the ChPT result is not a fit to the lattice data, it is fixed by a few input parameters as discussed in section \ref{sect:ImpactPrelim}. 

The fact that ChPT describes the data well for all $t'$, even close to source and sink, is surprising and hard to understand. We argued that the time separations $t-t'$ and $t'$ need to be sufficiently large such that pion physics dominates the 3-pt function, and we naively expected a minimal separation of about 1 fm for both $t-t'$ and $t'$. Therefore, for a source-sink separation as small as $t\approx 1$ fm we may expect ChPT to describe the 3-pt function and the ratio for $t'$ close to $t/2$, if at all. In other words we may have expected to roughly reproduce the slope the data exhibits in the middle of fig.\ \ref{fig:2}. The good agreement might simply be accidental, this possibility cannot be ruled out without a more detailed study with additional data.

As for the induced pseudoscalar form factor data we can analytically remove the LO $N\pi$ contribution from the data. In analogy to \pref{pacsdatacorrected} we compute corrected data according to
\begin{equation}\label{R4corrected}
R_4^{\rm corr}(Q^2,t,t') \equiv \frac{R_4(Q^2,t,t')}{1+ X_{4}(Q^2,t,t')}.
\end{equation}
Provided the NLO corrections and excited-state effects other than $N\pi$ are small we have (cf.\ \pref{NpiConttot})
\begin{equation}\label{R4expectation}
R_4^{\rm corr}(Q^2,t,t') \approx \Pi_4(Q^2)\,.
\end{equation}
In practice we can make use of our earlier observation that the tree diagram contribution $Z_4^{\rm tree}$ dominates $X_{4}$ in \pref{pacsdatacorrected}. The corrected data is shown by the black data symbols in fig.\ \ref{fig:2}. As expected from the good agreement between the original data and the ChPT result the corrected data is essentially a constant as a function of $t'$, i.e.\ the data fulfill \pref{R4expectation}. Note that applying the correction formula is problematic near the $t'$ values where the ratio develops a zero. There deviations between the lattice data and the approximate LO ChPT results become amplified, as can be seen in fig.\ \ref{fig:2} for $t'-t/2\approx 0.15$ fm. 

As repeatedly said, the $N\pi$ contribution is really the dominant part in the ratio $R_4$, and one has to go to much larger source-sink separations in order to significantly suppress it. Figure \ref{fig:2} also shows the ChPT results for twice and three times the original source-sink separation, i.e.\ for $t\approx 2.1$fm (red dotted line) and $t\approx 3.2$fm (blue dotted line). Apparently, the slope of the curves goes to zero very slowly. 

The reservations we expressed at the end of the last section apply here as well. More data at more and larger source-sink separations are necessary to quantitatively corroborate our findings here. Still, the LO ChPT result for the $N\pi$ contribution accommodates qualitatively the features seen in lattice data for the ratio $R_4$.\footnote{Ref.\ \cite{Bali:2018qus} proposes the use of the projected axial vector current 
\begin{equation}\label{Aperp}
A_{\mu}^{\perp} = \left(g_{\mu\nu} - \frac{\overline{p}_{\mu}\overline{p}_{\nu}}{\overline{p}^2}\right) A^{\nu}
\end{equation}
as a method to construct combinations of correlation functions that suffer less from excited-state contaminations (we follow eq.\ (23) in Ref.\ \cite{Bali:2018qus} and use the Minkowski space notation in \pref{Aperp}.)
As linear combinations  we know the  $N\pi$ contribution in the ratios $R_{\mu}^{\perp}$ to LO ChPT from our results for the $R_{\mu}$. In practice each ratio $R_{\mu}^{\perp}$ involves the $N\pi$ state contribution of all four $R_{\mu}$, also those that stem from loop diagrams. For those we have no reason to believe that a source-sink separation of 1 fm is large enough for ChPT to apply, thus we refrained from comparing lattice data for $R_{\mu}^{\perp}$ with the LO ChPT results.}

%========================
\section{Concluding remarks}\label{sect:concl}
%========================

We have seen that the dominant $N\pi$-state contribution to the two axial form factors is of different origin. In case of $\GP(Q^2)$ it stems from a single pion carrying a spatial momentum $\vec{q}$ associated with the momentum transfer $Q^2$. To $\GA(Q^2)$, on the other hand, a whole tower of $N\pi$ states contributes. Diagrammatically spoken, the loop diagrams in fig.\ \ref{fig:Npidiags} contribute to $\GA(Q^2)$, while the tree diagrams contribute to $\GP(Q^2)$. We have argued that this difference most likely implies different minimal source-sink separations to apply the ChPT results. In case of $\GA(Q^2)$ we still need comparably large source-sink separations of about 2 fm or larger to sufficiently suppress the high-momentum $N\pi$ states. Since these states do not contribute to $\GP(Q^2)$ much smaller separations seem to be accessible in this case.

Even with this reasoning in mind the comparison of the ChPT results with numerical lattice data  in section \ref{sec:confr} works unexpectedly well. It suggests that the discrepancy between lattice results and experimental data and the ppd model is dominantly causes by the $N\pi$ state-contamination. We stress again that this conclusion needs to be consolidated by comparisons with more lattice data. Still, the procedure suggested in section \ref{sec:confr} to analytically remove the anticipated $N\pi$-state contamination seems promising.

Here we only studied the 3-pt functions of the axial vector current. It is straightforward to do the analogous calculation with the pseudoscalar density and to calculate the $N\pi$-state contribution to the pseudoscalar form factor $\GP(Q^2)$. Making use of the partially conserved axial vector current (PCAC) relation the three form factors are related. However, this relation was found to be significantly violated by the lattice estimates for the form factors \cite{Rajan:2017lxk,Jang:2018lup,Ishikawa:2018rew,Bali:2018qus}. The reason for this violation is not fully understood, but with the ChPT results for the $N\pi$-state contribution to all three form factors one can explicitly check what r\^{o}le the two-particle $N\pi$ states play here. 

\vspace{4ex}
\noindent {\bf Acknowledgments}
\vspace{2ex}

At various stages of this project I have benefitted from discussions with G. Bali, S.~Collins, E.~Epelbaum, J.~Green, M.~Gruber, R.~Gupta, Y.~Kuramashi, K.-L.~Liu, H.~Meyer, T.~Schulz, R.~Sommer, P.~Wein, T.~Wurm and T.~Yamazaki.
I thank C.~Alexandrou, J.~Green, T.~Wurm for sending me their lattice data. 
This work was supported by the German Research Foundation (DFG), Grant ID BA 3494/2-1.
\vspace{3ex}

\noindent{\bf Appendix}

%========================
\begin{appendix}
%========================
\section{Summary of the Feynman rules}\label{appFeynmanRules}
We employ the covariant formulation of baryon ChPT \cite{Gasser:1987rb,Becher:1999he}, and our calculations are done to LO in the chiral expansion. To that order the chiral effective Lagrangian consists of two parts only, ${\cal L}_{\rm eff}={\cal L}_{N\pi}^{(1)} + {\cal L}_{\pi\pi}^{(2)}$. Expanding this Lagrangian in powers of pion fields and keeping interaction terms with one pion field only we obtain
\begin{eqnarray}
\label{Leff}
{\cal L}_{\rm eff} &=& \overline{\Psi} \Big(\gamma_{\mu}\partial_{\mu} +M_N \Big)\Psi +\frac{1}{2}\pi^a \Big(- \partial_{\mu}\partial_{\mu} + M_{\pi}^2 \Big)\pi^a + \frac{ig_A}{2f}\overline{\Psi}\gamma_{\mu}\gamma_5\sigma^a \Psi \, \partial_{\mu} \pi^a\,.
\end{eqnarray}
The  nucleon fields $\Psi=(p,n)^T$ and $\overline{\Psi}=(\overline{p},\overline{n})$ 
contain the proton and the neutron fields $p$ and $n$.  $M_{\pi}$ denotes the pion mass, while $M_N$, $g_A$ and $f$ are the chiral limit values of the nucleon mass, the axial charge and the pion decay constant. 

The interaction term in \pref{Leff} leads to the well known nucleon-pion interaction vertex proportional to the axial charge. A factor $i$ appears here because we work in euclidean space-time. From the terms quadratic in the fields one reads off the nucleon and pion propagators. We find the time-momentum representation for the propagators convenient. In that representation the pion propagator reads
\begin{eqnarray}
G^{ab}(x,y)& = &  \delta^{ab}L^{-3}\sum_{\vec{p}} \frac{1}{2 E_{\pi}} e^{i\vec{p}(\vec{x}-\vec{y})} e^{-E_{\pi} |x_0 - y_0|}\,,\label{scalprop}
\end{eqnarray}
with  the pion energy given by $E_{\pi} =\sqrt{\vec{p}^2 +M_{\pi}^2}$. The nucleon propagator $S^{ab}_{\alpha\beta}(x,y)$ is given by
\begin{eqnarray}
S_{\alpha\beta}^{ab}(x,y)& = &  \delta^{ab} L^{-3}\sum_{\vec{p}} \frac{Z_{p,\alpha\beta}^{\pm}}{2E_N} e^{i\vec{p}(\vec{x}-\vec{y})} e^{-E_N |x_0 - y_0|}\,.
\end{eqnarray} 
$a,b$ and $\alpha,\beta$ refer to the isospin and Dirac indices, respectively. The factor $Z^{\pm}_{\vec{p}}$ in the nucleon propagator (spinor indices suppressed) is defined as
\begin{equation}
Z_{\vec{p}}^{\pm}=-i\vec{p}\cdot\vec{\gamma} \pm E_N \gamma_4+M_N\,, 
\end{equation}
where the $+$ ($-$) sign applies to $x_0 > y_0$ ($x_0 < y_0$), and the nucleon energy is given by $E_{N,\vec{p}}=\sqrt{|\vec{p}|^2 +M_N^2}$. 
The sum in both propagators runs over the discrete spatial momenta that are compatible with periodic boundary conditions imposed on the finite spatial volume, i.e.\ 
$\vec{p}=2\pi \vec{n}/L$  with $\vec{n}$ having integer-valued components.

For the computation of the 3-pt function we need the expression for the  axial vector current.
It can be obtained from the known effective Lagrangian in the presence of an external source field for the axial vector current \cite{Gasser:1987rb}, and is found to be given by 
\begin{eqnarray}
A_{\mu}^a & = & g_A  \overline{\Psi}\gamma_{\mu}\gamma_5 \sigma^a\Psi -\frac{1}{f}\epsilon^{abc} \pi^b \overline{\Psi}\gamma_{\mu}\sigma^c\Psi - 2i f\partial_{\mu} \pi^a\,.\label{DefAxial}
\end{eqnarray}
The first two terms on the right hand side stem from ${\cal L}_{N\pi}^{(1)}$, the remaining one from ${\cal L}_{\pi\pi}^{(2)}$. The same expression has already been used in Refs.\ \cite{Bar:2016uoj}. 

Finally, the expressions for the nucleon interpolating fields in ChPT have been derived in Ref.\ \cite{Wein:2011ix}. To LO and up to one power in pion fields one finds
\begin{eqnarray}\label{Neffexp}
N(x)& = & \tilde{\alpha} \left(\Psi(x) + \frac{i}{2f} \pi^a(x)\sigma^a \gamma_5\Psi(x)\right)\,,\\
\overline{N}(0) & = & \tilde{\beta}^* \left(\overline{\Psi}(0) + \frac{i}{2f}\overline{\Psi}(0)\gamma_5\sigma^a\pi^a(0) \right)
\end{eqnarray}
These are the effective fields for the standard nucleon interpolating fields composed of three quarks without derivatives \cite{Ioffe:1981kw,Espriu:1983hu}. The interpolating fields not necessarily need to be point-like, but can also be constructed from `smeared' quark fields. These operators map to the same chiral expressions provided the smearing procedure is compatible with chiral symmetry and the `smearing radius' is small compared to the Compton wavelength of the pion. In that case smeared interpolating fields are mapped onto point like fields in ChPT just like their pointlike counterparts at the quark level  \cite{Bar:2013ora,Bar:2015zwa}. The expressions differ only by the LECs $\tilde{\alpha},\tilde{\beta}$. If the same interpolating fields are used at both source and sink we find $\tilde{\alpha}=\tilde{\beta}$. 

%===========
\section{LO results for the correction coefficients}\label{app:corrcoeff}
%===========
The correction coefficients $B_{k}^{\rm corr}(\vec{q},\vec{p}),\tilde{B}_{k}^{\rm corr}(\vec{q},\vec{p})$ and $C_{k}^{\rm corr}(\vec{q},\vec{p})$ are introduced in eqs.\ \pref{def:Bk} and \pref{def:B4}. The explicit results read:
\begin{eqnarray}
B_{k}^{\rm corr}(\vec{q},\vec{p}) &=&B_k^{\infty}(\vec{q},\vec{p})\left(  \frac{M_{\pi}^2 - pq}{\Epip^2}-\frac{1}{2 g_A}\right) - \frac{
(g_A^2-2) \Epiq^2  (p_3 q_k + p_k q_3)}{\Epip^2 \,q_k q_3}
\\
B_{3}^{\rm corr}(\vec{q},\vec{p}) &=&B_3^{\infty}(\vec{q},\vec{p}) \left( \frac{M_{\pi}^2}{\Epip^2} -\frac{1}{2 g_A} \right) - \frac{\Epiq^2 \{2 p^2 + (4 - g_A^2) p_3 q_3\}}{
 \Epip^2 (\Epiq^2 - q_3^2)}\nn\\
 & & + 
 \frac{\Epiq^2 \{\Epip^2 (2 - g_A^2) + g_A^2 (p^2 - 2 p_3^2\} pq}{
 \Epip^4 (\Epiq^2 - q_3^2)}\\
B_{4}^{\rm corr}(\vec{q},\vec{p})&=&- B_4^{\infty}(\vec{q},\vec{p})\left(\frac{4 - 4 g_A + 3 g_A^3}{8 g_A} + \frac{(2 + g_A^2) p^2+g_A^2 pq}{
    4 \Epip^2} \right)-g_A^2 \frac{\Epiq^2p^2}{\Epip^2 M_{\pi}^2} 
 \end{eqnarray}

\begin{eqnarray}
\tilde{B}_{k}^{\rm corr}(\vec{q},\vec{p}) &=&B_k^{\infty}(\vec{q},\vec{p})\left(  \frac{M_{\pi}^2 - pq}{\Epip^2}-\frac{1}{2 g_A}\right) - \frac{
\Epiq^2 \Big((g_A^2+g_A-2) p_3 q_k + (g_A^2-2g_A+2) p_k q_3)\Big)}{\Epip^2 \,q_k q_3}\nn\\
&&\\
\tilde{B}_{3}^{\rm corr}(\vec{q},\vec{p}) &=&B_3^{\infty}(\vec{q},\vec{p}) \left( \frac{M_{\pi}^2}{\Epip^2} -\frac{1}{2 g_A} \right) - \frac{\Epiq^2 \{2 p^2 + g_A p_3 q_3\}}{
 \Epip^2 (\Epiq^2 - q_3^2)} \\
 & & + 
 \frac{\Epiq^2 \{\Epip^2 (-2 -g_A+7 g_A^2) - g_A^2 (5p^2 + 2 p_3^2)\} pq}{
 \Epip^4 (\Epiq^2 - q_3^2)}\\
\tilde{B}_{4}^{\rm corr}(\vec{q},\vec{p})&=&- B_4^{\infty}(\vec{q},\vec{p})\left(\frac{4 - 4 g_A + 3 g_A^3}{8 g_A}+ \frac{(2 + g_A^2) p^2+(4+g_A^2 )pq}{
    4 \Epip^2} \right) \nn\\
& &-\frac{2\Epiq^2\Big(\Epip^2(4g_A-2)-g_A^3p^2\Big)}{g_A \Epip^2 M_{\pi}^2} 
\end{eqnarray}

\begin{eqnarray}
C_{k}^{\rm corr}(\vec{q},\vec{p}) &=&
C_k^{\infty} (\vec{q},\vec{p})\left(\frac{g_A-1}{g_A} - \frac{p^2 + pq}{\Epip^2}\right) + 
 \frac{ g_A (2 g_A-1) \Epiq^2(p_3 q_k + p_k q_3)}{2 \Epip^2\, q_k q_3}\\
C_{3}^{\rm corr}(\vec{q},\vec{p}) &=& C_3^{\infty} (\vec{q},\vec{p}) \left( \frac{M_{\pi}^2}{\Epip^2}-\frac{1}{g_A}\right)  - \frac{2g_A^2\Epiq^2   (p^2 - 2 p_3^2) pq}{2 \Epip^4 (\Epiq^2 - q_3^2)}\nn\\
& &  + \frac{
     g_A(2 g_A-1) \Epiq^2(pq - 2 p_3 q_3)}{2 \Epip^2 (\Epiq^2 - q_3^2)}\\
 C_{4}^{\rm corr}(\vec{q},\vec{p})    &= &-g_A^2 \frac{\Epiq^2 \left\{ 2p_3 (p^2 + pq) -  p^2 q_3\right\}}{\Epip^2 M_{\pi}^2 q_3} 
\end{eqnarray}
Note that all the coefficients given above are real. Together with the purely imaginary SN contribution the ratios $R_{\mu}$ are purely imaginary. However, some of the loop diagrams do lead to imaginary parts in the correction coefficients, implying a non-vanishing real part for the ratios with spatial indices. Therefore, the expressions given above apply to the ratios if we consider, in slight contrast to \pref{DefRatio}, the definition
\begin{equation}\label{DefRatio2}
R_{k}(\vec{q},t,t') =\frac{{\rm Im}\, C_{3,{k}}(\vec{q},t,t')}{C_2(0,t)}\sqrt{\frac{C_2(\vec{q},t-t')}{C_2(0,t-t')}\frac{C_2(\vec{0},t)}{C_2(\vec{q},t)}\frac{C_2(\vec{0},t')}{C_2(\vec{q},t')}}\,,
\end{equation}
i.e.\ if we simply drop the real part in the ratios with $k=1,2,3$.

\end{appendix}
%========================

\end{document}